\documentclass[amsfonts,amsmath,aps,nofootinbib,reprint,superscriptaddress]{revtex4-2}
\usepackage[dvipdfmx]{graphicx}
\usepackage{bm,latexsym,amsmath,amssymb,amsfonts} 
\DeclareMathAlphabet{\mathpzc}{OT1}{pzc}{m}{it}
\usepackage[normalem]{ulem}
\usepackage[breaklinks=true,colorlinks=true]{hyperref}
\usepackage{xcolor}
\usepackage{wasysym}
\usepackage[mathscr]{eucal}
\usepackage{multirow}
\usepackage{enumitem}
\usepackage{tabularray}
\usepackage{tablefootnote}
\usepackage{longtable}
\usepackage{booktabs}
\usepackage{lipsum}
\DeclareMathOperator{\sech}{sech}
\newcommand{\orcid}[1]{\href{https://orcid.org/#1}{\includegraphics[width=8pt]{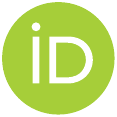}}}
\hypersetup{linkcolor=red, citecolor=blue, urlcolor=blue}

\begin{document}

\def\Journal#1#2#3{\href{https://doi.org/#3}{{\it #1} #2}}
\def\arXiv#1#2{\href{https://arxiv.org/abs/#1}{arXiv:#1 [#2]}}
        
\title{Horizonless star based on regular black hole with finite radius\\and its observational signatures}

\author{M.~F.~Fauzi\orcid{0009-0005-3380-6005}}
\thanks{Corresponding author.}
\email{muhammad.fahmi31@ui.ac.id}
\affiliation{Departemen Fisika, FMIPA, Universitas Indonesia, Depok 16424, Indonesia}

\author{B.~N.~Jayawiguna\orcid{0000-0001-5920-8701}}
\email{nugrahabyon312@gmail.com}
\affiliation{High Energy Physics Theory Group, Department of Physics, Faculty of Science, Chulalongkorn University, Bangkok 10330, Thailand.}
\affiliation{Department of Physics, University of Virginia, 382 McCormick Road, Charlottesville, Virginia 22904-4714, USA.}

\author{H.~S.~Ramadhan\orcid{0000-0003-0727-738X}}
\email{hramad@sci.ui.ac.id}
\affiliation{Departemen Fisika, FMIPA, Universitas Indonesia, Depok 16424, Indonesia}

\author{A.~Sulaksono\orcid{0000-0002-1493-5013}}
\email{anto.sulaksono@sci.ui.ac.id}
\affiliation{Departemen Fisika, FMIPA, Universitas Indonesia, Depok 16424, Indonesia}

\begin{abstract}
The horizonless configuration of regular black holes has recently attracted attention as a model for ultracompact stars. In this paper, we propose a new class of regular black hole models sourced by a de Sitter vacuum with a finite radius. We focus on studying its horizonless configuration, which is modified into an anisotropic gravastar by proposing an ansatz of equation of states. We confirm that an anisotropic gravastar approaching horizon formation must violate the dominant energy condition. We also found that the proposed object has an effectively similar structure as a frozen star on the time geometry at the extremal configuration. From the proposed model, we investigate the photon geodesics inside the object and predict the optical appearance of the object surrounded by a thin accretion disk. Our imaging results indicate that, assuming light does not interact with the object's interior, its optical appearance differs from that of a thin-shell gravastar. ``Chaotic" photon ring merges for $x>x_{m}$, where $x_{m}$ represents the minimum value required for the photon sphere to exist. In addition to its optical appearance, we investigate the axial gravitational perturbations emitted by this horizonless star.  Notably, echo trains are found to exist for $x>x_{m}$, as determined by numerically solving the time-dependent Regge–Wheeler equation. By comparing the echo time with the GW170817 observation, we find that a frequency of 72 Hz can be achieved, albeit at the cost of requiring a relatively high value of $\ell$.
\end{abstract}

\maketitle

\section{Introduction}
Pathologies arising from Einstein's theory of general relativity (GR) have long posed challenges in fundamental physics, with black hole (BH) singularities being a prime example. According to standard classical GR, two types of `infinities' arise in and around a BH: the event horizon and the spacetime singularity, associated with infinite time for an infalling-observer and infinite curvature, respectively. Recent discussion related to these issues can be seen in Ref.~\cite{Carballo-Rubio:2025fnc}. The well-known weak cosmic censorship conjecture (WCCC) suggests that these two infinities are connected, with the spacetime singularity always hidden behind the horizon, making it unobservable from the outside world \cite{Penrose:1969pc}.

In an attempt to resolve the singularity problem, numerous nonsingular models of BHs have been proposed in the literature, commonly referred to as regular BHs (RBHs)~\cite{Lan:2023cvz}. They are sourced by different types of physics mechanism that may arise in extreme curvature or density, such as nonlinear electrodynamics~\cite{Ayon-Beato:2000mjt,Ayon-Beato:1998hmi,Bronnikov:2000vy,Burinskii:2002pz,Bronnikov:2021uta}, de Sitter vacuum~\cite{Hayward:2005gi,Cadoni:2022vsn,Cadoni:2022chn}, or quantum gravity~\cite{Nicolini:2023hub,Ashtekar:2023cod,Eichhorn:2022bgu}. The absence of a singularity in RBHs renders the WCCC irrelevant, and in most cases, it is possible to obtain an RBH solution without a horizon. A widely known quantity that controls this solution is $\alpha$, which generally represents the ratio between the BH's mass and the new physical quantity that regulates the BH~\cite{Cadoni:2022chn,Carballo-Rubio:2022nuj}.

Alongside the proposals for RBHs, several models of BH mimickers have been introduced as alternatives to BHs (for a review, see Ref.~\cite{Cardoso:2019rvt}). A BH mimicker can be simply defined as an object that does not possess any horizons but shares a similar structure and properties with BHs. These mimickers might take the form of traversable wormholes or ultracompact objects, such as gravastar.

The Gravastar, short for \textit{gravitational vacuum condensate star}, was proposed by Mazur and Mottola~\cite{Mazur:2001fv}. It suggests that extreme gravitational collapse and high density could trigger a phase transition, resulting in an object with a de Sitter interior that has a repulsive equation of state (EoS), $p = -\epsilon$. This object is held up by a thin shell of matter at its surface, with an EoS of $p = \epsilon$, allowing for a smooth transition between the de Sitter interior and the Schwarzschild vacuum exterior. These transitions are typically discontinuous, with a radial pressure jump at the boundaries between the interior, thin shell, and exterior, leading to a Dirac-delta transverse pressure. However, there are also models of gravastar which has continuous radial pressure transition, dubbed as the \textit{anisotropic gravastar}, introduced by Cattoen \textit{et al.}~\cite{Cattoen:2005he} and later realized by DeBenedictis \textit{et al.}~\cite{DeBenedictis:2005vp} who obtain the numerical solutions. Cattoen \textit{et al.}~\cite{Cattoen:2005he} argued that gravastars must violate the dominant energy conditions, particularly near the point where the local compactness approaches the formation of a horizon. Numerous gravastar models have been developed within various theoretical frameworks~\cite{Adler:2022fqu, Sanjay:2024mkg, Ghosh:2023wps, Moti:2021vck, Bhatti:2021xqi, Beltracchi:2021lez, Horvat:2008ch, Chan:2011ayt}, including those featuring continuous radial pressure transitions~\cite{Cattoen:2005he, DeBenedictis:2005vp, Jampolski:2023xwh, Fauzi:2024nta}.

It has been shown that a BH mimicker can potentially be related to a regular black hole (RBH), with the possibility of transitioning between the two through various mechanisms~\cite{Carballo-Rubio:2025fnc, Carballo-Rubio:2022nuj}. One proposed scenario for this transition is from a de Sitter vacuum RBH to an anisotropic gravastar, where they share similarities in their core, which is filled with an anisotropic de Sitter vacuum and has an EoS of $p = -\epsilon$~\cite{Carballo-Rubio:2022nuj, Fauzi:2024nta}. To our knowledge, Carballo-Rubio \textit{et al.}~\cite{Carballo-Rubio:2022nuj} present the first proposal for this scenario, suggesting that a change in mass or a new scale could interpolate into a horizonless star solution with a structure similar to that of an anisotropic gravastar. However, the radial pressure in their model is negative throughout, as it assumes no modification to the de Sitter EoS, though the transverse pressure does have a region with a positive value. Cattoen \textit{et al.}~\cite{Cattoen:2005he} argued that positive radial pressure must build up in the outer regions of the star, as negative radial pressure leads to repulsive gravity, thus supporting the transition between the de Sitter interior and Schwarzschild vacuum exterior. 

Recently, three of the authors~\cite{Fauzi:2024nta} proposed a horizonless solution that includes a region of positive radial pressure in the star's atmosphere, making it more suitable as an anisotropic gravastar solution. However, this model introduces a discontinuous cut-off in the energy density, resulting in a discontinuity in the radial metric component. Hence, to address the discontinuity problem, this work aims to develop a generalized framework for constructing an anisotropic gravastar based on RBH spacetime with finite radii, allowing for a smooth interpolation between each solution, and to investigate some of its key properties.

The existence of these theoretical models can only be confirmed through astrophysical observations. One conventional method is imaging the object with radio telescopes, as has been done for BHs through the Event Horizon Telescope's (EHT) observations of M87*~\cite{EventHorizonTelescope:2019dse,EventHorizonTelescope:2024dhe} and Sagittarius A* BHs~\cite{EventHorizonTelescope:2022wkp}. Extensive investigations of these images have been conducted, most of which constrain the parameters from beyond GR theories. The key quantity revealed by these observations is the shadow radius: the radius of the dark patch appearing in the image of the BHs, which represents the magnified radius of the horizon. However, assuming that light follows the corresponding spacetime geodesic without interaction with its surrounding, it has been shown that in a horizonless spacetime containing photon sphere(s), the shadow feature is absent and is replaced by ring-like secondary images~\cite{Fauzi:2024nta,Eichhorn:2022fcl,Chen:2024ibc,Rosa:2023qcv,Rosa:2023hfm,Rosa:2024eva,Guerrero:2022msp,Rosa:2024bqv}\textemdash here dubbed as the \textit{photon rings}. With current observational instruments, these photon rings may not be resolved due to low resolution, but they may be revealed in future observations with higher-resolution instruments, potentially by the next-generation EHT~\cite{Eichhorn:2022fcl}.

The properties of photon ring behavior could serve to distinguish horizonless ultracompact objects from other types of astrophysical objects. For example, in the semi-classical horizonless compact object model discussed by Chen and Yokokura~\cite{Chen:2024ibc}, and the anisotropic gravastar model examined by Fauzi \textit{et al.}~\cite{Fauzi:2024nta}, multiple thin photon rings are observed near the Schwarzschild photon ring, which are referred to here as \textit{chaotic photon rings} due to their erratic behavior. In contrast, several studies by Rosa \textit{et al.} on thin-shell gravastars~\cite{Rosa:2024bqv}, constant density stars (CDS)~\cite{Rosa:2023hfm}, and boson stars~\cite{Rosa:2023qcv} do not exhibit these chaotic rings for objects with $R < 3M$. These chaotic photon rings may represent a distinctive feature of certain spacetime geometries, offering a promising method for identifying corresponding objects in future observations.

Another critical phenomenological aspect relevant to astrophysical observations is the study of non-radial oscillations and Gravitational Wave (GW) events. The progression from the initial attempts to detect gravitational waves, in the first observation of a BH merger (GW150914) \cite{LIGOScientific:2016aoc}, the binary neutron star (NS) inspiral (GW170817) \cite{LIGOScientific:2017vwq}, and the detection of two compact object coalescences (GW200105 and GW200115) \cite{LIGOScientific:2021qlt}, has confirmed the existence of gravitational waves. These observations now allow for the study of GWs both experimentally and analytically. The emission of gravitational wave signals by compact objects depends on several factors, including the mass, radius, moment of inertia, and the degree to which the object can deform, often characterized by quadrupole deformation. The compactness of an object, defined as the ratio between its mass and radius, plays a key role in its classification \cite{Cardoso:2019rvt}. The BH regime is reached when the object’s radius is exactly twice its mass, $R=2M$. A second classification, the ultracompact object region, corresponds to objects capable of emitting gravitational wave signals. This region is expected to exist due to the presence of a photon sphere—an unstable region where massless particles can orbit at specific radii. The latter aspect is believed to play a significant role in the emission of gravitational waves. From a theoretical standpoint, the study of these phenomena can be framed within the context of perturbation theory.

In Newtonian theory, the adiabatic perturbation is governed by a fourth-order linear differential system that couples the Newtonian potential to the perturbation variable \cite{ferrari}. Typically, a time-independent oscillatory form of $e^{i\omega t }$, where $\omega$ is a constant frequency, is assumed. The system is solved by imposing boundary conditions: (i) regularity at the center of the star, and (ii) vanishing pressure at the star’s surface. These conditions lead to a discrete set of real values for the frequency, denoted by $\omega_{n}$. In the relativistic case, this framework is viewed as a generalization of the Newtonian theory. The governing equation decomposes into two types of perturbations. The first, the polar (even) mode, corresponds to tidal oscillations and has a Newtonian analog. The second, the axial (odd) mode, induces stationary rotation of the star and does not have a Newtonian counterpart, making it a purely relativistic phenomenon. The axial perturbation, described by a master equation, can be reduced to a one-dimensional Schrödinger-like equation with a corresponding effective potential. However, the gravitational waves associated with these axial perturbations, particularly those exhibiting echo patterns, are fully characterized by this effective potential. The Schrödinger-like equation reveals that the characteristic frequencies of these waves are complex, with the imaginary part of the frequency being physically interpreted as the damping of mechanical energy, leading to an exponential decay in the wave signal. In this study, we focus specifically on the axial perturbations of RBHs, modeled according to our constructed horizonless star model.

The gravitational perturbation theory has been extensively applied to various EoS in the literature. In \cite{ferrari}, the author investigates both polar and axial modes within the context of the constant density star model and demonstrates that, in the axial mode, resonance can occur if the star exceeds the ultracompact object regime. In this case, the well of the effective potential traps the gravitational waves. Similar characteristics are observed for exotic EoSs, as shown in \cite{Urbano:2018nrs}. The CDS is considered an ultra-stiff EoS, as it allows for a nonzero constant density throughout the interior of the star. However, in \cite{Urbano:2018nrs}, it is noted that the pressure diverges when the radius of the star with this EoS reaches $R=2.52M$. Furthermore, the CDS model is unphysical due to the infinite speed of sound within the star. In contrast to the CDS, the Tolman VII (TVII) model \cite{Tolman:1939jz,Lattimer:2000nx,Raghoonundun:2015wga,Raghoonundun:2016cun,Hensh:2019rtb,Stuchlik:2021coc,Stuchlik:2021vdd,Posada:2021zxk,Posada:2023bnm} offers a more realistic EoS. The TVII solution represents one of the few analytic solutions to the Einstein field equations, with the energy density profile being nonzero at the center of the star and monotonically decreasing towards the surface. Outside the star, the metric reduces to the Schwarzschild spacetime. For TVII, the scaling axial and polar GW modes are studied in Ref.~\cite{Tsui:2004qd} whereas the quasinormal modes are studied in Ref.~\cite{Neary:2001ai,Neary:2001yg}. To further improve the description of more realistic neutron stars, Jiang and Yagi \cite{Jiang:2019vmf} introduced a modified Tolman-VII (MTVII) model, which incorporates an additional quartic term into the energy density profile, with a free parameter $\alpha$, to better capture the physics of neutron stars. Numerical solutions for this model are explored in \cite{Posada:2022lij}. Additionally, the gravitational perturbations for both the TVII and MTVII models in modified gravity have been studied in \cite{Jayawiguna:2022ftj,Jayawiguna:2023vvw}.

The structure of this paper is as follows. In Sec.~\ref{sec. I field eq}, we discuss our initial framework of GR and RBHs. Next, in Sec.~\ref{sec. II modified hayward} we propose a RBH model based on Hayward RBH with a finite radius. In Sec.~\ref{sec. III anisotropic grav}, we construct an effective EoS to construct the anisotropic gravastar, and present the results on pressure profile and energy conditions. In Sec.~\ref{sec. IV optical appearance}, we investigate the geodesics inside the object and its optical appearance when it surrounded by thin accretion disk. In Sec.~\ref{sec. V gravitational perturb}, we study the other observational signature, that is, the GW signals, including the quasinormal modes and the GW echoes. Finally, in Sec.~\ref{sec. summary} we summarize and discuss our main results.

\subsection*{\textit{Executive Summary}}

Here, we summarize the main results for convenience. 

\begin{enumerate}
	\item We propose a type of de Sitter RBH model based on the Hayward RBH with a finite radius $R$, achieved by adding a 'Tolman-like' term to the energy density profile,
\begin{align}
    \bar{\epsilon}(y) &= \frac{3}{8\pi} \frac{\bar{\kappa}\alpha}{(y^3 + 1)^2}\left[1 - \left(\frac{y}{\bar{R}}\right)^n\right],\notag\\ 
    \bar{m}(y) &= \frac{\bar{\kappa}\alpha}{2} y^3 \left[\frac{1}{y^3 + 1} - \bar{\mathcal{H}}(y) \right],\notag\\
    \bar{\mathcal{H}}(y) &= \frac{3\left(y/\bar{R}\right)^n {}_2F_1\left(2,1+\frac{n}{3};2+\frac{n}{3};-y^3\right)}{3+n},
\end{align}
where $\bar{\epsilon} = \epsilon \ell^2$ is the energy density, $\bar{m} = m / \ell$ is the mass profile, also known as the Misner-Sharp (MS) mass, and $\alpha = 2\mathcal{M} / \ell$, with $\mathcal{M}$ being the total mass or the Arnowitt-Deser-Misner (ADM) mass, and $\ell$ is a parameter that regulates the BH. From the corresponding energy density, we also propose an ansatz for the EoS that deviates from the de Sitter EoS in the horizonless configuration,
\begin{equation}
    \bar{p}(y) = - \bar{\epsilon}(y) \left[1 - F(y) \Theta(x-1)\right],
\end{equation}
where $F(y)$ is an arbitrary function satisfying the required conditions, and $\Theta(x-1)$ is a cut-off function with $x = \alpha / \alpha_c$. As an example, we choose a function that effectively satisfies the anisotropic gravastar requirements,
\begin{align}
    F(y) &= \mathcal{T}(y) \left[1 + a\left(\frac{\bar{\epsilon}(y)}{\bar{\epsilon}_0}\right)^{\gamma-1}\right],\notag\\
    \mathcal{T}(y) &= \frac{1}{2}\left[1 + \tanh\left(\frac{y-x\omega}{x\sigma_t}\right)\right].
\end{align}
with $a$, $\gamma$, $\omega$, and $\sigma_t$ are some parameters that can be determined or constrained.

\begin{figure}[htbp!]
\centering 
\includegraphics[width=.49\textwidth]{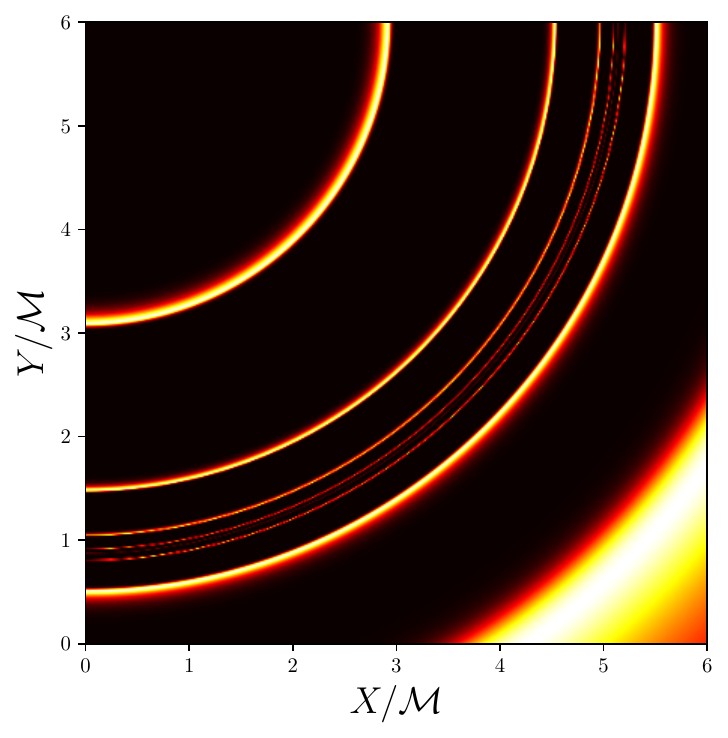}
\caption{A close up image of our star model surrounded by ISCO accretion disk with $x=0.9$ and $\bar{R}=1.1$.}
\label{fig. exsecutive summary - chaotic minor photon rings}
\end{figure}

\item We predict the observational signatures of the object. We first generate the appearance of the object surrounded by a thin accretion disk by performing a ray tracing procedure. Our results show that, in the presence of photon spheres and assuming the object is surrounded by a thin accretion disk extending from the ISCO radius outward, the generated images consist of chaotic minor photon rings between the first two major photon rings (see Fig.~\ref{fig. exsecutive summary - chaotic minor photon rings}). These feature do not appear in the images of the thin shell gravastar model (see Ref.~\cite{Rosa:2024bqv} or Fig.~\ref{fig. axial thin shell}).

\item We extend our results to the observation of gravitational echoes using axial perturbation theory. In our analysis, a key parameter, denoted as $x_{m}$, emerges, which represents the minimum value for a compact object to exhibit photon sphere properties. This behavior is shown in the context of the effective potential and the corresponding gravitational echo plot. Additionally, the wave propagates within the quasinormal mode spectrum.  We obtain that as the potential well becomes deeper, the compact object enables the higher trapped mode. The estimated frequency from the echo time calculations shows that our horizonless star model remains far from the expected results based on existing echo observational data.

\end{enumerate}

\section{Field equation and regular black hole}
\label{sec. I field eq}
\subsection{Perfect fluids and energy conditions}
With the static and spherically symmetric spacetime ansatz in form of
\begin{equation}
    ds^2 = -e^\nu dt^2 + e^\lambda dr^2 + r^2 d\Omega^2,
    \label{eq. SSS exponent}
\end{equation}
solution to the Einstein field equation with energy-momentum tensor of an anisotropic perfect fluid satisfying
\begin{equation}
    T^{\mu\nu} = (\epsilon + p_t) u^\mu u^\nu + p_t g^{\mu\nu} + (p-p_t)s^\mu s^\nu,
\end{equation}
are given by
\begin{align}
	e^{-\lambda}\left(\frac{\nu'}{r}+\frac{1}{r^2}\right)-\frac{1}{r^2} &= 8\pi p, \label{eq. 1st Guv = Tuv}\\
	e^{-\lambda}\left(\frac{1}{2}\nu'' - \frac{1}{4}\lambda'\nu'+\frac{1}{4}(\nu')^2+\frac{\nu'-\lambda'}{2r}\right) &= 8\pi p_t, \label{eq. 2nd Guv = Tuv}\\
	e^{-\lambda}\left(\frac{\lambda'}{r} - \frac{1}{r^2}\right) + \frac{1}{r^2} &= 8\pi \epsilon,
	\label{eq. 3rd Guv = Tuv}
\end{align}
where $\epsilon$, $p$, and $p_t$ are the energy density, radial pressure, and transverse pressure of the perfect fluid, respectively. The Eq.~\eqref{eq. 3rd Guv = Tuv} results in
\begin{equation}
    e^{-\lambda(r)} = 1-\frac{2m(r)}{r}, \qquad m(r)\equiv4\pi \int_{0}^{r} \epsilon(\tilde{r}) \tilde{r}^2 d\tilde{r}.
    \label{eq. EFE mass profile}
\end{equation}
and the TOV equation reads
\begin{equation}
    \frac{dp}{dr} = -\left[\epsilon(r) + p(r)\right] \frac{m(r) + 4\pi p(r) r^3}{r\left[r-2m(r)\right]} + 2 \frac{p_t(r) - p(r)}{r}.
    \label{eq. TOV anisotropic}
\end{equation}

With anisotropic perfect fluid energy-momentum tensor, Hawking and Ellis \cite{Hawking:1973uf} mentioned several constraints known as energy conditions. In general, there are three main types: the {\it strong energy condition} (SEC), the {\it null energy condition} (NEC) which underlies the more general weak energy condition (WEC), and the {\it dominant energy condition} (DEC). They are given by
\begin{align}
    \epsilon + p + 2p_t &\equiv SEC \geq 0, & \label{eq. SEC}\\
    \epsilon + p &\equiv WEC_1 \geq 0, & \epsilon + p_t &\equiv WEC_2\geq 0,\label{eq. NEC}\\
    |p| &\equiv DEC_1 \leq \epsilon, & |p_t| &\equiv DEC_2 \leq \epsilon.\label{eq. DEC}
\end{align}
All of these energy conditions are also followed by the requirement of non negative energy density, that is $\epsilon > 0$. The physical implementation of these energy conditions are given as follows~\cite{Hawking:1973uf,Curiel:2014zba}:
\begin{enumerate}
    \item \textbf{SEC} implies that gravity acts locally attractive to an observer following timelike geodesics.
    \item \textbf{WEC} implies that energy density is non-negative as measured by any observer.
    \item \textbf{DEC} is basically the same as WEC, with an additional requirement that the pressure could not exceed the energy density. This comes from the fact that the speed of sound must be less than the speed of light. In natural units, this implies $dp/d\epsilon<1$, and violation of DEC would guarantee that the speed of sound is superluminal. It also applies to the transverse pressure $p_t$.
\end{enumerate}

In some cases, these three energy conditions might be violated, especially the SEC. Cosmological implication of the energy conditions can be found in Ref.~\cite{Visser:1999de}. Violation of SEC itself implies `repulsive' gravity, which is the known nature of dark energy. In the case of matter, the satisfaction of energy condition\textemdash specifically the SEC and WEC\textemdash leads to a classification of types of matter~\cite{Chan:2008ui,Chan:2011ayt}; \textit{normal matter}, \textit{dark energy}, and \textit{phantom energy}. Normal matter, as the name suggest, would satisfy every energy conditions, while dark energy only violates the SEC but satisfies WEC. Phantom energy, on the other hand, violates at least one of the WEC, and it can either satisfy (attractive) or violates (repulsive) the SEC. The conditions and classifications are given in Table.~\ref{tab. matter classification}.
\begin{table}[htbp!]
    \centering
    \begin{tabular}{lcccccc}
        \hline \hline
        \multicolumn{1}{c}{Matter} & \hspace{0.2cm} & \multicolumn{1}{c}{$SEC$} & \hspace{0.2cm} & \multicolumn{1}{c}{$WEC_1$} & \hspace{0.2cm} & \multicolumn{1}{c}{$WEC_2$}\\\hline
        Normal matter & & \checkmark & & \checkmark & & \checkmark\\
        Dark energy & & $\times$ & & \checkmark & & \checkmark\\
        Repulsive phantom energy && $\times$ && $\times$ &&\checkmark\\
        && $\times$ && \checkmark && $\times$\\
        && $\times$ && $\times$ && $\times$\\
        Attractive phantom energy && \checkmark && \checkmark && $\times$\\
        && \checkmark && $\times$ && \checkmark
        \\ \hline
    \end{tabular}
    \caption{Summarized classification of matter based on the satisfaction of energy conditions.}
    \label{tab. matter classification}
\end{table}

\subsection{Hayward-type regular black hole}
Here we take Hayward model as the representation of RBH models, because in general~\cite{Cadoni:2022chn}, RBH models share generic features. However, Hayward model is the a simplest model that has stable cores~\cite{Bonnano2021}. Considering a RBH energy density in form of~\cite{Hayward:2005gi,Cadoni:2022chn}
\begin{equation}
    \epsilon(r) = \frac{3}{4\pi} \frac{\mathcal{M}\ell^3}{\left(r^3 + \ell^3\right)^2},
\end{equation}
and imposing the de Sitter EoS ($p=-\epsilon$) yields to a model of Hayward RBH, with the metric component written as
\begin{equation}
    e^\nu = e^{-\lambda} = 1-\frac{2m(r)}{r},\qquad m(r) = \frac{\mathcal{M}r^3}{r^3 + \ell^3}.
\end{equation}
Here, $\mathcal{M}$ is the Arnowitt-Deser-Misner (ADM) mass of the BH, and $\ell$ can be considered as a new physics scale that regulates the spacetime. Hayward RBH model is classified as the de Sitter RBH for the fact that in the limit $r\to0$, it behaves as a de Sitter spacetime with $m(r\to0)\propto r^2/\ell^2$.

It is useful to define a quantity $\alpha$,
\begin{equation}
    \alpha = \frac{2\mathcal{M}}{\ell},
    \label{eq. alpha def}
\end{equation}
where there exists a critical value $\alpha_c$ which determines the characteristic of the metric solution:
\begin{enumerate}
    \item if $\alpha > \alpha_c$, the solution contains two horizons,
    \item if $\alpha = \alpha_c$, the solution contains one extremal horizon, and
    \item if $\alpha < \alpha_c$, the solution is horizonless.
\end{enumerate}
The value of $\alpha_c$ can be determined by demanding $e^{-\lambda}=\partial_r e^{-\lambda} = 0$. In the case of Hayward RBH, it is known that $\alpha_c = 3/\sqrt[3]{4}$ and the extremal horizon radius is located at $r_c = \sqrt[2]{3}\ell$ or, in terms of the ADM mass, $r_c = 4\mathcal{M}/3$~\cite{Cadoni:2022chn}.

\section{Modified Hayward metric with finite radius}
\label{sec. II modified hayward}
We consider the energy density profile in form of
\begin{equation}
    \epsilon(r) =
    \begin{cases}
        \frac{3}{4\pi} \frac{M\ell^3}{\left(r^3 + \ell^3\right)^2}\left[1 - \left(\frac{r}{R}\right)^n\right], & 0\leq r \leq R\\
        0, & r>R
    \end{cases}
    \label{eq. density profile M}
\end{equation}
where $R$ is interpreted as the \textit{surface radius} of the object. Introducing the aditional `Tolman-like' $1-(r/R)^n$ terms allows us to cut off the energy density smoothly at the surface, which is a widely convenient way to define a surface radius. This modification is a phenomenological approach to include a surface term, allowing the star solution to smoothly interpolate from the RBH solution\textemdash in contrast to the solution proposed in our previous study~\cite{Fauzi:2024nta}. Now, it should be noted that $M$ is a constant and \textit{does not} correspond to the total mass of the object. 

Solving the field equation of Eq.~\eqref{eq. EFE mass profile}, one obtain
\begin{align}
    m(r) &= 
    \begin{cases}
        Mr^3\left[\frac{1}{r^3 + \ell^3} - \mathcal{H}(r)\right], & 0\leq r \leq R\\
        m(R), & r>R
    \end{cases}\notag\\
    \mathcal{H}(r) &= \frac{3\left(\frac{r}{R}\right)^n {}_2F_1\left(2,\frac{n}{3}+1,\frac{n}{3}+2,-\frac{r^3}{\ell^3}\right)}{\ell^3(3+n)},
    \label{eq. mass profile M}
\end{align}
where $\mathcal{H}(r)$ is a new term as a result of the finite radius correction. In the limit of large $R$, $\mathcal{H}(r)|_{R\to\infty}=0$ and the original Hayward's mass profile is retrieved. It is obvious that $m(R)$ is the total mass of the object. The total mass is closely related with the ADM mass $\mathcal{M}$, where
\begin{equation}
    \mathcal{M} = m(R) = MR^3 \left[\frac{1}{\ell^3+R^3}- \mathcal{H}(R)\right].
\end{equation}
We can introduce a new constant,
\begin{equation}
    \kappa = \frac{M}{\mathcal{M}}\equiv \left[R^3 \left(\frac{1}{\ell^3+R^3}-\mathcal{H}(R)\right)\right]^{-1},
\end{equation}
so that $M$ in Eqs.~\eqref{eq. density profile M} and~\eqref{eq. mass profile M} can be written in terms of total mass,
\begin{align}
    \epsilon(r) &=
    \begin{cases}
        \frac{3}{4\pi} \frac{\kappa\mathcal{M}\ell^3}{\left(r^3 + \ell^3\right)^2}\left[1 - \left(\frac{r}{R}\right)^n\right], & 0\leq r \leq R\\
        0, & r>R
    \end{cases}\notag\\
     m(r) &= 
     \begin{cases}
         \kappa\mathcal{M}r^3\left[\frac{1}{r^3 + \ell^3} -\mathcal{H}(r)\right], & 0\leq r \leq R\\
         \mathcal{M}, & r>R.
     \end{cases}
     \label{eq. modified hayward energy and mass}
\end{align}
Now we have two spacetime region
\begin{align}
    e^{-\lambda(r)} = 
    \begin{cases}
        1 - 2\kappa\mathcal{M}r^2\left[\frac{1}{r^3 + \ell^3} - \mathcal{H}(r)\right], & 0\leq r \leq R\\
        1 - \frac{2\mathcal{M}}{r}, & r>R.
    \end{cases}
\end{align}
The solution at the boundary $R$ is continuous up to the first derivative. Hence, the modification to the spacetime from the Schwarzschild only extend from $r=0$ to $r=R$, and it interpolates smoothly with the Schwarzschild exterior. Additionally, one can achieve the Schwarzschild BH solution by simply setting $\ell\to0$ with any arbitrary value of $R$, resulting in $\mathcal{H}\to0$ and $m(r)\to\mathcal{M}$ in every region of the spacetime.

To simplify the calculation, we define several dimensionless variables
\begin{gather}
y\equiv \frac{r}{\ell},\qquad \bar{R} \equiv \frac{R}{\ell},\notag\\
\bar{\epsilon} \equiv \epsilon\ell^2, \qquad \bar{p} \equiv p\ell^2, \qquad \bar{p}_t \equiv p_t\ell^2, \qquad \bar{m} \equiv \frac{m}{\ell},
\end{gather}
so that the dimensionless energy density and Misner-Sharp mass becomes a function of $y$
\begin{align}
    \bar{\epsilon}(y) &= \frac{3}{8\pi} \frac{\bar{\kappa}\alpha}{(y^3 + 1)^2}\left[1 - \left(\frac{y}{\Bar{R}}\right)^n\right],\notag\\ 
    \bar{m}(y) &= \frac{\bar{\kappa}\alpha}{2} y^3 \left[\frac{1}{y^3 + 1} - \bar{\mathcal{H}}(y) \right],\notag\\
    \bar{\mathcal{H}}(y) &= \frac{3\left(y/\bar{R}\right)^n {}_2F_1\left(2,1+\frac{n}{3};2+\frac{n}{3};-y^3\right)}{3+n},
    \label{eq. modified mass and energy density y}
\end{align}
where
\begin{equation}
    \bar{\kappa} = \left[\bar{R}^3 \left(\frac{1}{\bar{R}^3+1}-\bar{\mathcal{H}}(\bar{R})\right)\right]^{-1}.
\end{equation}
The TOV equation now becomes
\begin{equation}
    \frac{d\bar{p}}{dy} = -\left(\bar{\epsilon} + \bar{p}\right) \frac{\bar{m} + 4\pi\bar{p}y^3}{y\left(y-2\bar{m}\right)} + 2\frac{\bar{p}_t - \bar{p}}{y}.
\end{equation}

\begin{figure}[htbp!]
\centering 
\includegraphics[width=.49\textwidth]{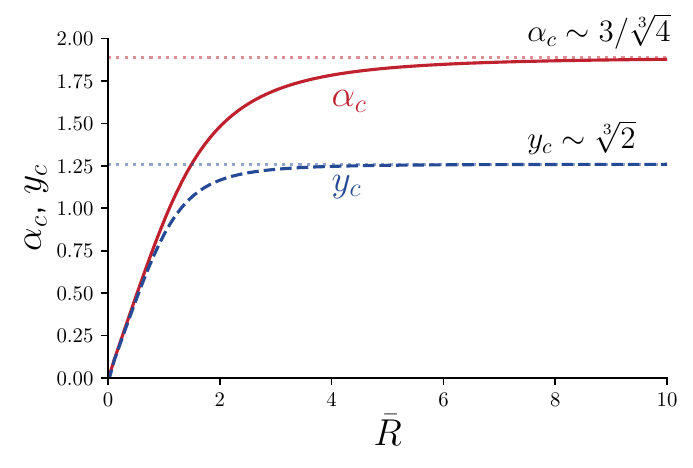}
\hfill
\caption{Values of $\alpha_c$ and $y_c$ for given value of $\bar{R}$, with $n=3$. The $\alpha_c$ and $y_c$ approaches the standard Hayward value of $\alpha_c = 3/\sqrt[3]{4}$ and $y_c = \sqrt[3]{2}$ as $\bar{R}\to \infty$.}
\label{fig. ac and rh of R}
\end{figure}

Since the metric solution is complicated, we use numerical calculation to determine the $\alpha_c$ value and its extremal horizon location $y_c$, which, as expected, has dependent on $\bar{R}$. The result is shown in Fig.~\ref{fig. ac and rh of R}. We see that the $\alpha_c$ value is modified in the small $\bar{R}$ regime, and it approaches the original Hayward value as $\bar{R} \to \infty$. For our modification to be significant, we shall focus on a surface radius comparable to $\ell$, with $\bar{R} \geq 1$, as we expect the surface radius to be greater than the new scale $\ell$. 

However, the value of $\ell$ might be extremely small if one expects the modification to appear near the Planck scale $\ell_p$, so that $\ell \sim \ell_p$~(see, for instance, Refs.~\cite{Bonanno:2000ep,Platania:2019kyx,Eichhorn:2022bbn}). On the other hand, several studies also indicates the possibility that the modification of $\ell$ might be relevant in horizon or cosmological scale (for a brief overview, see Ref.~\cite{Cadoni:2022chn}). Hence, we consider the possibility that this scale can be arbitrarily large, making it relevant to astrophysical phenomena and potentially observable through astronomical-scale observations.

\section{Anisotropic gravastar as horizonless star solution}
\label{sec. III anisotropic grav}

This study suggest that the horizonless configuration of our RBH model be replaced by an anisotropic gravastar, following the previous work in Ref.~\cite{Fauzi:2024nta}. Our approach is based on the idea that, in the absence of a horizon, the repulsive core could cause positive pressure matter to accumulate outside, forming a region of positive pressure that acts as a transition layer between the Schwarzschild vacuum outside and the de Sitter core within~\cite{Cattoen:2005he}. When horizons are present, it is possible to assume that the horizon itself could catalyze the infalling positive pressure matter, which pottentially create negative pressure around the horizon extending to the radius of the innermost stable circular orbit (ISCO).

Based on Ref.~\cite{Cattoen:2005he}, anisotropic gravastar solution demands following conditions;
\begin{enumerate}
    \item the de Sitter core requires $\bar{p}(0) = \bar{p}_t(0) = -\bar{\epsilon}(0)$,
    \item there must be two radial point where the $p$ vanishes: at $y=y_0$ and $y=\bar{R}$; $y_0$ is the transition point between negative pressure of de Sitter vacuum and positive pressure of the atmosphere,
    \item $\partial_y\bar{p}(0) =0$ to maintain the regularity at the center.
\end{enumerate}
\begin{figure}[htbp!]
\centering 
\includegraphics[width=.49\textwidth]{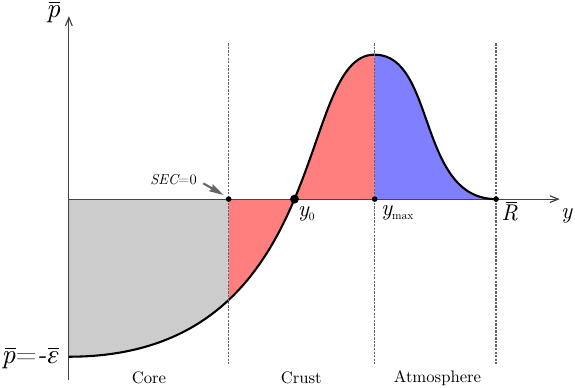}
\caption{Qualitative sketch of anisotropic gravastar's  pressure profile and its region.}
\label{fig. pressure illustration}
\end{figure}
Some restrictions were also given regarding the energy conditions. The WEC must be satisfied throughout the entire region of the star, while the SEC is indeed violated near the core of the star. However, there are no restriction for the DEC, since it is expected to be violated especially near the formation of horizons, which has been shown in  Refs.~\cite{DeBenedictis:2005vp,Jampolski:2023xwh,Fauzi:2024nta} that the DEC is indeed violated. Based on the requirement and energy conditions, anisotropic gravastar is divided by three main interior region, ilustrated in Fig.~\ref{fig. pressure illustration};
\begin{enumerate}
	\item \textit{the core}, located from the star center $y=0$ to a point where the SEC satisfied,
	\item \textit{the crust}, located from the outer side of the core to the point of maximum positive radial pressure $y=y_{m}$ where $d\bar{p}/dy=0$,
	\item \textit{the atmosphere}, located outside the crust to the star surface $\bar{R}$.
\end{enumerate}
Here, we observe that the anisotropic gravastar has a dark energy core, with what is considered `normal matter' on the crust through the surface. However, the increasing pressure in the crust region would imply that the speed of sound, \( v_s^2 = d\bar{p}/d\bar{\epsilon} \), becomes imaginary. Therefore, we conclude that the actual normal matter applies only in the atmosphere region, where we demand real and subluminal speed of sound, i.e. $0\leq v_s^2 \leq 1$.

\subsection{Effective equation of state}
Now we demand an effective EoS that satisfy the requirement for anisotropic gravastar, while transition into de Sitter $\bar{p}=-\bar{\epsilon}$ at the horizonful configuration limit. It is useful to define another dimensionless quantity
\begin{equation}
    x=\frac{\alpha}{\alpha_c},
\end{equation} so that $x<1$ for horizonless configuration. We will only consider $y<\bar{R}$ in what follows, since $\bar{p}(y>\bar{R}) = \bar{\epsilon}(y>\bar{R}) = 0$.

For the transition from the de Sitter EoS RBH to the anisotropic gravastar EoS, we simply demand
\begin{equation}
    \bar{p}=
        \begin{cases}
            -\bar{\epsilon}, & x \gtrsim 1 \qquad \text{(Black hole)}\\
            -\bar{\epsilon}[1 - F(y)], & x < 1 \qquad \text{(Gravastar)}
        \end{cases}
\end{equation}
so that below the extremal limit $x < 1$, the EoS shifts away from the de Sitter form. We introduce $F(y)$ as an arbitrary function representing the modification to the de Sitter EoS, which gives rise to the buildup of positive pressure outside the core. To make the transition smoother, we propose the following ansatz for the EoS:
\begin{equation}
    \bar{p}(y) = - \bar{\epsilon}(y) \left[1 - F(y) \Theta(x - 1)\right],
\end{equation}
where $\Theta(x - 1)$ is a cutoff function satisfying
\begin{equation}
    \Theta(x - 1) \approx 
    \begin{cases}
        0, & x \gtrsim 1\\
        1, & x < 1
    \end{cases}
\end{equation}
A typical smooth function that satisfies this requirement is the sigmoid jump function,
\begin{equation}
    \Theta(x - 1) = \frac{1}{1 + \exp\left[{(x - 1)/\sigma_s}\right]},
\end{equation}
where $\sigma_s$ determines the `smoothness' of the transition. In the limit $x \to 1$, the function takes the value $1/2$. This non-zero cutoff may cause the extremal configuration's EoS to deviate from the de Sitter form, leading to a deformation of the spacetime such that it is no longer time-radially symmetric ($e^{\nu} \neq e^{-\lambda}$).

Let us now recall the necessary conditions for the anisotropic gravastar solution mentioned earlier. Within the ansatz, the first condition require that $F(0) = 0$, so the function expansion near $y\to0$ cannot have a constant term. The former requirement of the second condition implies that
\begin{equation}
    F(y_0)\Theta(x-1) = 1
    \label{eq. F r0 = 1},
\end{equation}
while the latter is automatically satisfied as $\bar{\epsilon}(\bar{R})=0$. It should be noted that, generally, $1-F(y)\Theta(x-1)$ should only have one root. Taking the derivative of the pressure ansatz, we have
\begin{align}
    \bar{p}'(y) = &-\bar{\epsilon}'(y) + \bar{\epsilon}(y)F'(y)\Theta(x-1) \notag\\
    &+ \bar{\epsilon}'(y) F(y) \Theta(x-1),
\end{align}
where the prime denotes the derivative with respect to $y$. Hence, $F'(0) = 0$ so that it must be at least of order two in $y$. The WEC gives a restriction such that $F(y)>0$ everywhere. If we also demand the dominant energy condition of the radial pressure, we require
\begin{equation}
    \left|1-F(y)\Theta(x-1)\right| \leq 1
    \label{eq. dominant energy},
\end{equation}
which implies that $0<F(y)\Theta(x-1)<2$. Moreover, since we are interested in the horizonless spacetime, we consider $x<1$ so that we can neglect the $\Theta(x-1)$ term in the following discussion.

Within this EoS ansatz, the transverse pressure can be obtained by solving the TOV equation of Eq.~\eqref{eq. TOV anisotropic}, which results in
\begin{align}
    \bar{p}_t(y) =  &F^2(y) H(y) + F(y) Q(y)\notag\\
    & + F'(y) \frac{\bar{\epsilon}(y) y}{2} - \frac{\bar{\epsilon}'(y) y}{2} - \bar{\epsilon}(y),
    \label{eq. transverse pressure}
\end{align}
where
\begin{align}
    H(y) &= \frac{4\pi \bar{\epsilon}^2(y) y^3}{2 \left[y-2\bar{m}(y)\right]},\notag\\
    Q(y) &= \bar{\epsilon}(y)\left(1 + \frac{\bar{m}(y)- 4\pi\bar{\epsilon}(y)y^3}{2\left[y-2\bar{m}(y)\right]}\right) + \frac{\bar{\epsilon}'(y)y}{2}.
\end{align}
We can clearly see the transverse pressure behavior at the surface as any terms that contains $\bar{\epsilon}(r)$ vanishes, so that we will only left with
\begin{equation}
    \bar{p}_t(\bar{R}) = \frac{\bar{R}\bar{\epsilon}'(\bar{R})}{2}\left[F(\bar{R}) - 1\right].
\end{equation}
Since $\bar{\epsilon}'(\bar{R}) < 0$, if one requires zero transverse pressure at the surface, then $F(\bar{R})$ must be unity. Relaxation of this condition introduces a new requirement: $0 < F(\bar{R})\leq 1$, so that the WEC is always satisfied on the surface. However, if $F(\bar{R})<1$, it can easily be seen that there will be two roots of $1-F(y)$ that are located in $y=y_0$ and $y_0<y<\bar{R}$ and violate the number of pressure transition points. Hence, we shall enforce $F(\bar{R}) = 1$.

Overall, for objects with non-zero $F(y)$ and local compactness close to forming a horizon, \textit{i.e.}, $2m(y)/y \to 1$, a significant increase in transverse pressure would occur at the location of maximum local compactness, potentially violating the DEC and confirming the claim of Ref.~\cite{Cattoen:2005he}. This can actually be avoided by setting $F(y) \propto 1 - 2m(y)/y$, multiplied by additional factors to enforce $F(0) = F'(0) = 0$. However, the additional factors would complicate satisfying the other requirements, so we will not consider this form in our discussion.

To conclude all the requirements, we summarize the required conditions of $F(y)$ in Table.~\ref{tab. F(r) condition}.

\begin{table}[htbp!]
    \centering
    \begin{tabular}{lclcl} \hline \hline
        \multicolumn{1}{c}{Requirement} & \hspace{0.2cm} & \multicolumn{1}{c}{Pressure Condition} & \hspace{0.2cm} & \multicolumn{1}{c}{$F(y)$ Condition}\\ \hline
        Central dS core & & $\bar{p}(0)=-\bar{\epsilon}(0)$ & & $F(0)=0$\\
        Regularity & & $\bar{p}'(0)=0$  & & $F'(0)=0$\\
        Pressure transition & & $\bar{p}(y_0) = 0$  & & $1-F(y_0)=0$ $^\dagger$\\
        WEC & & $\bar{\epsilon}(y) + \bar{p}(y) \geq 0$ & & $F(y) \geq 0$ \\
         & & $\bar{\epsilon}(\bar{R}) + \bar{p}_t(\bar{R}) \geq 0$ & & $F(\bar{R})= 1$\\
        DEC & & $|\bar{p}(y)|\leq \bar{\epsilon}(y)$ & & $0\leq F(y)\leq 2$\\
        \hline
    \end{tabular}
    \raggedright
    {\footnotesize $^\dagger$ One root at $0<y<\bar{R}$}
    \caption{Summarizing the requirement of $F(y)$ function.}
    \label{tab. F(r) condition}
\end{table}

\textbf{Other constraints from energy conditions}\textemdash We can ensure that more energy conditions are satisfied by adding several other constraints to the $F(y)$ function, but it might not be possible to determine them analytically. We get a deeper look at the transverse pressure $\bar{p}_t$, as the energy conditions of the radial pressure are easily (and have already been) obtained by the effective EoS itself. From Eq.~\eqref{eq. transverse pressure}, the WEC of Eq.~\eqref{eq. NEC} reads
\begin{equation}
    F^2(y) H(y) + F(y) Q(y) + F'(y) \frac{\bar{\epsilon}(y) y}{2} - \frac{\bar{\epsilon}'(y) y}{2} > 0.
\end{equation}
We can directly see two terms that potentially violate the WEC: $Q(y)$ and $F'(y)$. From the graphic analysis with our form of energy density, $Q(y)$ is always negative around the atmosphere due to the $\epsilon'(y)$ term. The $F'(y)$ term will be negative or zero at the atmosphere if one requires $\bar{p}_t(\bar{R})=0$, since there must exist a point $y_0$ such that $F(y_0)=1$.

We also demand another condition to the model: subliminal speed of sound on the atmosphere. The speed of sound reads
\begin{equation}
    v_s^2 = \frac{d\bar{p}}{d\bar{\epsilon}} = \frac{d\bar{p}}{dy} \left(\frac{d\bar{\epsilon}}{dy}\right)^{-1}.
\end{equation}
With our ansatz, the equation becomes
\begin{equation}
    v_s^2 =  \frac{\bar{\epsilon}(y) F'(y)}{\bar{\epsilon}'(y)} + F(y) - 1,
\end{equation}
and the speed of sound at the surface is $v_s^2 = F(\bar{R}) - 1$. The maximum speed of sound location can be found by finding the root of $d(v_s^2)/dy = 0$ at the atmosphere, and it yields to
\begin{gather}
     F'(y_{s})\left[2-\frac{\bar{\epsilon}(y_{s})\bar{\epsilon}''(y_{s})}{(\bar{\epsilon}'(y_{s}))^2}\right] + F''(y_{s}) \frac{\bar{\epsilon}(y_{s})}{\bar{\epsilon}'(y_{s})} = 0, \\ (y_m\leq y\leq\bar{R}) \notag
\end{gather}
where $y_s$ is the location of the maximum speed of sound point. Hence, $v_s^2 (y_s) \leq 1$ is required if we demand subluminal speed of sound at the atmosphere.

\subsection{Spacetime metric}

We decompose the metric in Eq.~\eqref{eq. SSS exponent} into
\begin{equation}
    ds^2 = -e^{\Phi}e^{-\lambda} dt^2 + e^{\lambda} dr^2 + r^2 d\Omega^2,
\end{equation}
where $\nu \equiv \Phi - \lambda$ is the \textit{shift} function. This approach is useful to investigate the consequences of the modification to the original metric. The Einstein field equation of Eq.~\eqref{eq. 1st Guv = Tuv} becomes
\begin{equation}
    e^{-\lambda}\left(\frac{\partial_r\Phi - \partial_r\lambda}{r} + \frac{1}{r^2}\right) - \frac{1}{r^2} = 8\pi p.
\end{equation}
Within our EoS ansatz and employing the dimensionless variables, we have
\begin{equation}
    \Phi' = \frac{8\pi y^2 \bar{\epsilon}(y)F(y)\Theta(x-1)}{y-2\bar{m}(y)},
    \label{eq. Phi'}
\end{equation}
The shift function $\Phi(y)$ can be evaluated analytically or numerically, depending on the form of $F(y)$. Alternatively, we can take the opposite approach, where we first define the desired $\Phi(y)$ and then determine the corresponding $F(y)$ function. Although this approach is easier, it is more challenging to find a solution that satisfies the anisotropic gravastar requirements.

This equation tells us that a slight deviation from the de Sitter EoS\textemdash that is, a non-zero $F(y)\Theta(x - 1)$\textemdash would deform the spacetime metric from the time-radial symmetric form, such that $e^{\nu} \neq e^{-\lambda}$. Since one requires $F(y) > 0$ to satisfy the WEC (cf. Table.~\ref{tab. F(r) condition}), the core region with non-zero $F(y)\Theta(x - 1)$ becomes redshifted. The redshift becomes more significant as the local compactness approaches the formation of a horizon ($m(y) \to y/2$).

As previously mentioned, the extremal configuration with $x = 1$ corresponds to a non-zero $\Theta(x - 1)$. It means that the interior of the object\textemdash especially the region below the critical point $y_c$\textemdash becomes infinitely redshifted, where $e^{\phi} \to 0$, effectively turning it into an object with similar structure as a frozen star on the $e^{\nu}$ component~\cite{Brustein:2021lnr,Brustein:2024yyc,Brustein:2023hic,Brustein:2024sah,Brustein:2023gea}.

\subsection{Effective $F(y)$ model}

A phenomenological form of $F(y)$ is already proposed in Ref.~\cite{Fauzi:2024nta}, which can be expressed as
\begin{equation}
    F(y) = g(y)\frac{\bar{R}}{\bar{m}(\bar{R})} \frac{\bar{m}(y)}{y},
\end{equation}
where $g(y)$ is an arbitrary function that satisfies several requirement (see the mentioned reference for the detail). Although it can effectively satisfy the anisotropic gravastar profile with varying $\bar{R}$, it faces several issues within our modified Hayward energy density and constant $\bar{R}$. The relative radial pressure profile $\bar{p}/\bar{\epsilon}_0$ does not depend on $x$, as
\begin{equation}
    \frac{\bar{R}}{\bar{m}(\bar{R})} \frac{\bar{m}(y)}{y} = \bar{\kappa} y^2 \bar{R} \left[\frac{1}{y^3 + 1} - \bar{\mathcal{H}}(y)\right].
\end{equation}
This behavior contradicts our expectations, as intuitively, we would expect the negative pressure region to increase as the star approaches a horizonful configuration\textemdash i.e., with increasing density or mass. This is also observed in the formation of dark energy stars, where the negative pressure region grows larger as more matter accumulates inside the star~\cite{Beltracchi:2018ait}. Therefore, here we propose an alternative $F(y)$ function that can satisfy all the required conditions.

There is a type of promising function that could effectively satisfy every $F(y)$'s requirement and can be easily adjusted/constrained: \textit{activation function}, such as the previously mentioned Sigmoid or the hyperbolic tangent (tanh). For example, lets consider an $F(y)$ function that is proportional to a hyperbolic tangent,
\begin{align}
    F(y) &= \mathcal{T}(y) \left[1 + a\left(\frac{\bar{\epsilon}(y)}{\bar{\epsilon}_0}\right)^{\gamma-1}\right],\notag\\
    \mathcal{T}(y) &= \frac{1}{2}\left[1 + \tanh\left(\frac{y-x\omega}{x\sigma_t}\right)\right].
    \label{eq. F(r) tanh epsilon}
\end{align}
With this function, one can achieve the polytropic-like EoS at the gravastar's atmosphere $\bar{p} \propto a\bar{\epsilon}^\gamma$, with $\gamma$ is related to the polytropic index and $a$ is a propotionality constant. It is inspired by the first anisotropic gravastar solution from Ref.~\cite{DeBenedictis:2005vp}, where the EoS approaches polytropic at the atmosphere. At the surface, this $F(y)$ behaves as 
\begin{equation}
F(\bar{R})\approx1+a\left(\frac{\bar{\epsilon}(\bar{R})}{\epsilon_0}\right)^{(\gamma-1)},
\end{equation}
so one require $\gamma>1$ to ensure the weak energy condition of transverse pressure at the surface. The $\omega$ and $\sigma_t$ will dominantly shape the transition, where the value of $\omega$ mostly determine the location of the transition point and $\sigma_t$ determine the smoothness of the transition. At the center $y=0$, the value of this function is
\begin{align}
    F(0) &= \frac{1}{2}\left(1 + a\right)\left[1 - \tanh\left(\frac{\omega}{\sigma_t}\right)\right],\notag\\
    F'(0) &= \frac{1 + a}{2x\sigma_t}\sech^2\left(\frac{\omega}{\sigma_t}\right),
    \label{eq. regularity condition center}
\end{align}
so that we require a proper value of $\omega$ and $\sigma_t$ to have central de Sitter core $p(0)\simeq -\epsilon_0$ and maintain its regularity, respectively, at least up to a desired order of magnitude. We also desire $F(\bar{R})=1$ at the surface, which require
\begin{equation}
    F(\bar{R}) = \frac{1}{2} \left[1+\tanh\left(\frac{\bar{R}-x\omega}{x\sigma_t}\right)\right] \approx 1,
    \label{eq. regularity condition surface}
\end{equation}
and can constraint the value of $\omega$ and $\sigma_t$ further.

There is a way to determine the value of $a$; by demanding that a positive value region appears as $\alpha<\alpha_c$. The idea is that, once the horizon is resolved, positive pressure matter begins to form near the surface. It can be achieved by ensuring that there is exactly only one root of $1-F(y)$ located at the surface, which is achieved by demanding
\begin{equation}
    F'(\bar{R})|_{x\to1} = 0.
\end{equation}
Within our ansatz and energy density, it will require
\begin{equation}
    a = \frac{\bar{R}(\bar{R}^3 + 1)^2 \sech^2\left(\frac{\bar{R}-\omega}{\sigma_t}\right)}{n \sigma_t \left[1 + \tanh\left(\frac{\bar{R}-\omega}{\sigma_t}\right)\right]}.
    \label{eq. determine a}
\end{equation}

Now one might wonder: \textit{why does $x$ term present inside the hyperbolic tangent?} This comes from the fact that we demand the size of the de Sitter core\textemdash or generally, the negative pressure region\textemdash gets larger with $x$ value. Decreasing $x$ term on the nominator shifts the transition zone towards the center, while $x$ term on the denominator maintain the regularity condition at the center. However, this approach would not results in all positive matter pressure as $x \ll 1$, which is beyond the scope of our discussion.

\subsubsection{Anisotropic gravastar solution}

In this section, we discuss the results and implementations of the energy condition constraints to the proposed EoS. We fix $n=3$ and $\sigma_s=10^{-3}$ for the sake of simplicity. Since $\omega$ significantly determines the transition point, it makes sense to instead adjust $\bar{\omega} = \omega/\bar{R}$, so that the transition point is relative to the surface. We first discuss the constraints on the choice of $\sigma_t$ and $\bar{\omega}$ based on the regularity and surface conditions. Using the determined value, we show the representative pressure profiles, analyze the energy conditions, and plot the metric components.  

\begin{figure}[htbp!]
\centering 
\includegraphics[width=0.48\textwidth]{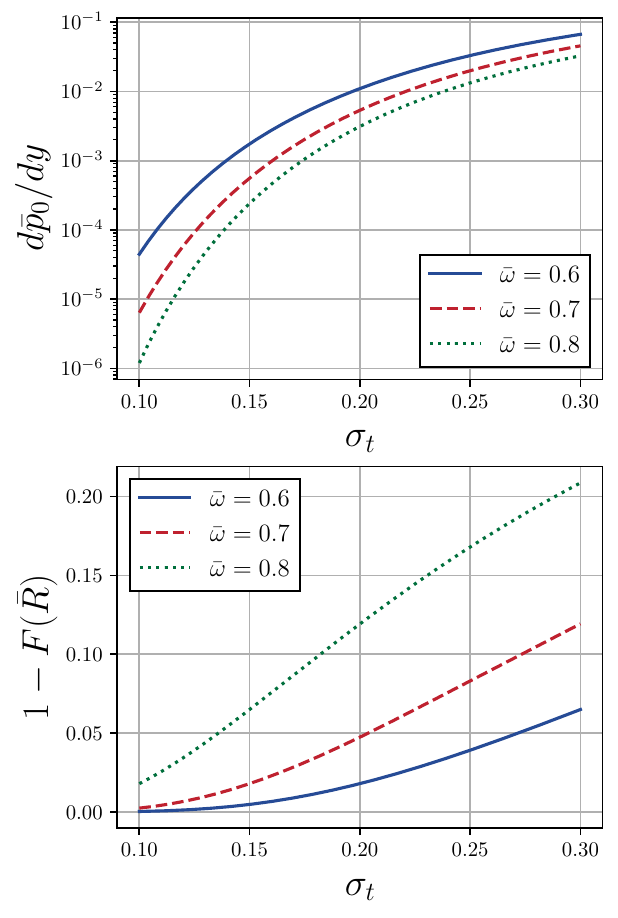}
\caption{Regularity ($d\bar{p}_0/dy$) and surface condition ($1-F(\bar{R})$) for values of $\sigma_t$ with representative value of $\bar{\omega}$. We set $\bar{R}=1$ and $x=1$.}
\label{fig. regularity condition}
\end{figure}

\textbf{Regularity \& surface conditions}\textemdash From Eqs.~\eqref{eq. regularity condition center} and \eqref{eq. regularity condition surface}, both conditions are sensitive to certain configurations. They are particularly vulnerable when $\bar{R}$ is small, and the surface condition is also sensitive to large values of $x$. Therefore, we analyze the two conditions with $\bar{R}=1$ and $x=1$, which represent the smallest and largest values of interest in our study, respectively. We show both conditions in Fig.~\ref{fig. regularity condition} for a range of $\sigma_t$ value with several representative $\bar{\omega}$.

It is expected that the regularity condition would easily be maintained by choosing arbitrarily small $\sigma_t$ value. However, this study considers a continuous pressure star model, and we demand smooth pressure transition which can be obtained by setting large $\sigma_t$ value where it potentially violates the regularity condition. Furthermore, the $\bar{\omega}$ value would affect the regularity and surface condition in opposite way. Hence, based on Fig.~\ref{fig. regularity condition}, we can choose a particular configuration of $\bar{\omega}$ and $\sigma_t$ values that favor both regularity and surface conditions, without having to sacrifice the transition smoothness. From our choice of $\bar{\omega}$ value, it is best to set $\bar{\omega}=0.7$ and $\sigma_t=0.15$ to maintain the regularity condition around the order of $10^{-4}$.

\begin{figure}[htbp!]
\centering 
\includegraphics[width=0.49\textwidth]{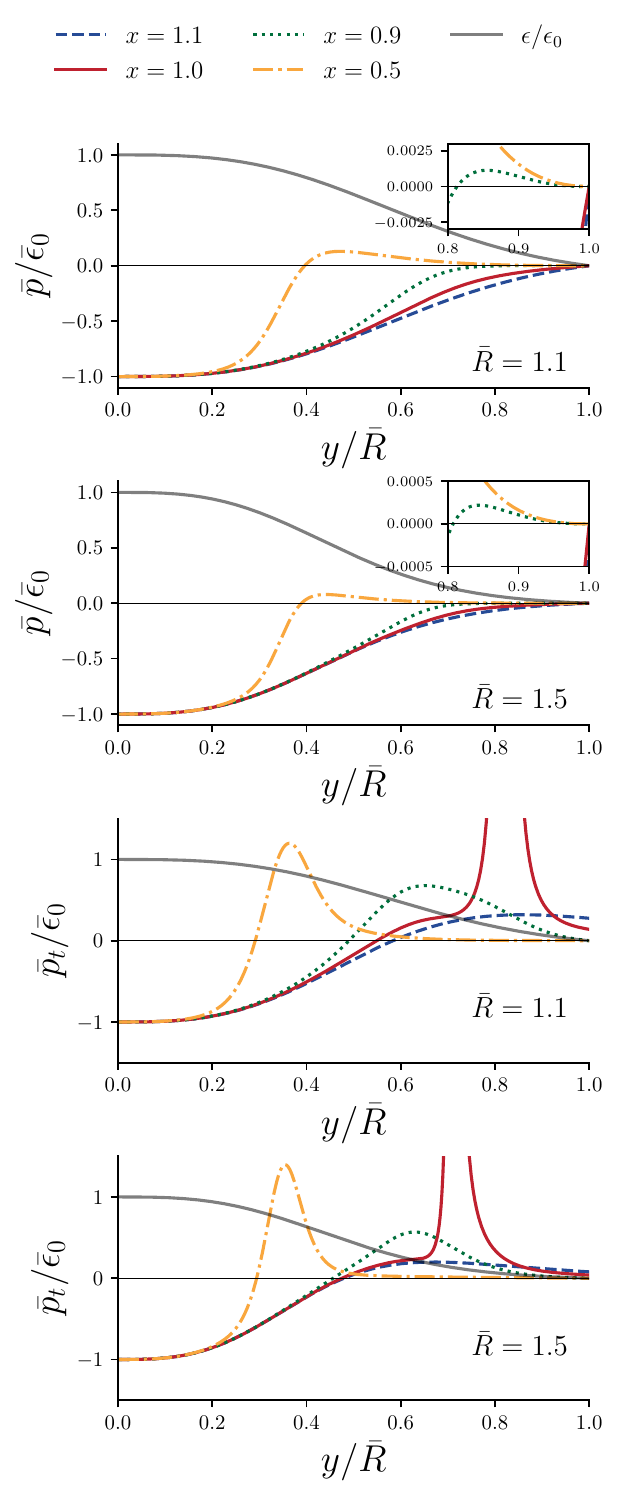}
\caption{Radial (left) and transverse pressure (right) profile alongside their corresponding energy density (gray solid line) for several configurations of $x$ with $\bar{R}=1.1$ and $\bar{R}=1.1$.}
\label{fig. pressure prof tanh}
\end{figure}

\begin{figure*}[htbp!]
\centering 
\includegraphics[width=0.9\textwidth]{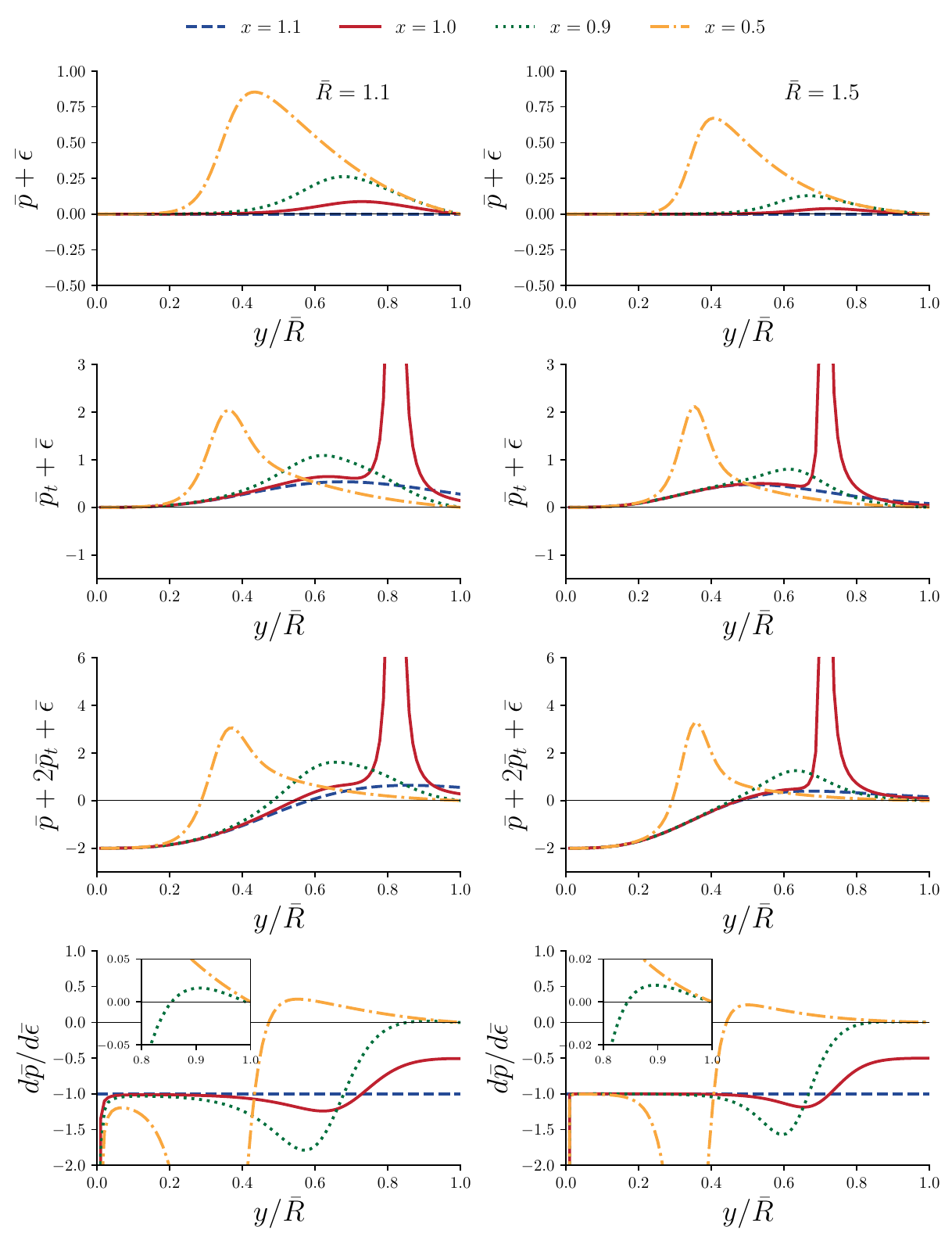}
\caption{Energy conditions and speed of sound representation for several configurations of $x$ with $\bar{R} = 1.1$ (top) and $\bar{R}=1.5$ (bottom).}
\label{fig. energy conditions tanh}
\end{figure*}

\textbf{Pressure profile and energy conditions}\textemdash Here we give a few examples to examine the pressure profile and energy conditions for a given set of configurations. It should be noted that to satisfy a physical sense, the radius of the star $\bar{R}$ should be higher than the new physics scale $\ell$, \textit{i.e.} $\bar{R}>1$. Hence, we choose $\bar{R}\in\{1.1, 1.5\}$ so that the radius is comparable to the new physics scale\textemdash thereby making our surface modification significant, as previously mentioned in Sec.~\ref{sec. II modified hayward}\textemdash and subsequently vary $x\in \{0.3, 0.5, 0.9\}$. The pressure profile and energy condition results are shown in Figs.~\ref{fig. pressure prof tanh} and \ref{fig. energy conditions tanh}, respectively.

The resulting radial pressure profile is aligned with our demand and the required condition of anisotropic gravastar: negative pressure, smoothly transition into positive pressure, no violations of dominant energy conditions, and the negative pressure region gets larger with higher $x$ value. With $x=0.9$, the positive pressure is barely visible since the configuration is close to becoming a BH with de Sitter EoS $\bar{p} = \bar{\epsilon}$, and it is a consequence from the choosen value of $a$ from Eq.~\eqref{eq. determine a}. 

For the transverse pressure, we shall compromise since there exists a region where the dominant energy condition is violated as shown by the region where $|\bar{p}_t|>\bar{\epsilon}$. This case is generally acceptable in anisotropic gravastar model, {\it e.g.} in Refs.~\cite{DeBenedictis:2005vp,Jampolski:2023xwh}, as Cattoen \textit{et al.}~\cite{Cattoen:2005he} suggests that anisotropic perfect fluid close to forming a horizon would violates DEC. In our case, it is excatly what happenes with large $x$, where the violation of DEC occurs when the local compactness $m(y)/y$ is close to forming a horizon ($2m(y)\sim y$), causing the $H(y)$ and the first term of $Q(y)$ dominate. However, DEC is also violated in the low local compactness, {\it i.e.}, with low $x$, mainly caused by the steep transition of $F(y)$, making the $F'(y)$ term dominate. It can be avoided by shifting the pressure transition inward (decreasing $\omega$) or employing smoother transitions (increasing $\sigma_t$), although it would be problematic for the regularity condition. Hence, a huge transverse pressure is needed for the object to support either its high compactness or the steep radial pressure transition.

From the energy condition analysis of Fig.~\ref{fig. energy conditions tanh}, we also see that our model satisfy every energy conditions requirement imposed by anisotropic gravastar model. The WEC are always positive for both radial and transverse pressure, indicating that the weak energy condition is satisfied throughout the entire region of the star. The SEC are also always positive at the outer region, while the negative region indicates the star's repulsive de Sitter core which is a feature of anisotropic gravastar. The speed of sound representation $d\bar{p}/d\bar{\epsilon}$ are also maintained below unity at the atmosphere, thus indicating that the matter is not superluminal. However, as a consequence, the higher and wider real and subluminal speed of sound at the atmosphere are followed by a larger imaginary peak from the radial pressure transition.

\begin{figure}[htbp!]
\centering 
\includegraphics[width=0.46\textwidth]{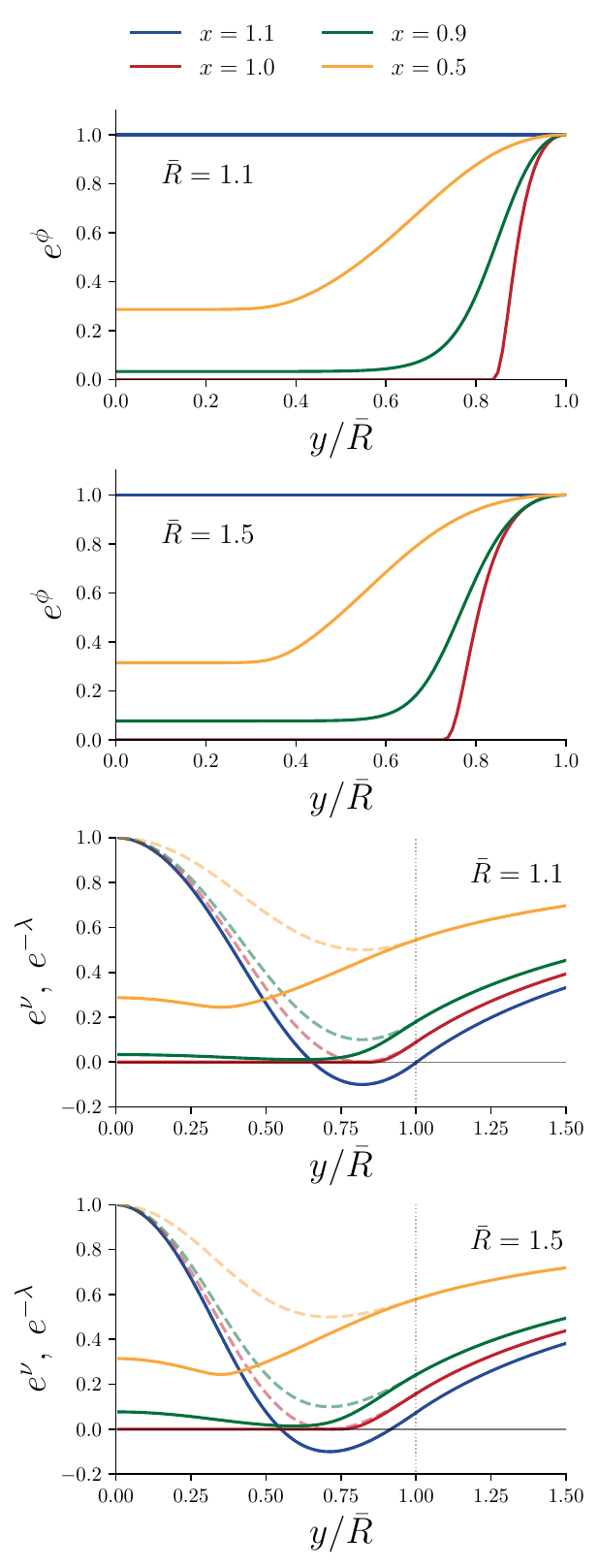}
\caption{(First and second row) Shift metric component $e^\phi$, and (third and fourth row) time component $e^\nu$ (solid) along with the radial component $e^{-\lambda}$ (dashed), with $\bar{R}=1.1$ and $\bar{R}=1.5$.}
\label{fig. metric component tanh}
\end{figure}

\textbf{Metric solutions}\textemdash The metric solutions can be obtained by integrating Eq.~\eqref{eq. Phi'} numerically. The integration is carried backward from the surface ($y=\bar{R}$) to the center ($y=0$), using the boundary condition $\Phi(\bar{R}) = 0$ to match the exterior Schwarzschild solution. We plot the shift, time, and radial metric component in Fig.~\ref{fig. metric component tanh}.

Since the central radial component $e^{-\lambda}$ always has a value of unity at the center, one can directly concludes that the shift function $e^\Phi$ plays role in the central gravitational redshift. It is observed that higher value of $x$ leads to higher redshifts (lower value of central $e^\Phi$), and it is significantly affected by the radius $\bar{R}$ in large $x$. However, as one can see from the last two rows of Fig.~\ref{fig. metric component tanh}, the maximum gravitational redshift is not located at the center, which is also the case for several other ultracompact objects \cite{Chen:2024ibc, Fauzi:2024nta}.

These results further support our argument for the formation of a frozen star structure in the extremal configuration, particularly evident in the behavior of the $e^{\nu}$ component. In the interior region, $e^{\nu}$ remains nearly constant and close to zero at the critical point $y_c$, indicating that time effectively freezes inside the object at extremality. This results in a structure resembling a frozen star, as described in Refs.~\cite{Brustein:2021lnr,Brustein:2024yyc,Brustein:2023hic,Brustein:2024sah,Brustein:2023gea}. When the configuration exceeds extremality, \textit{i.e.}, for $x > 1$, the shift function $e^{\phi}$ quickly transitions to a constant value of unity, becoming a usual de Sitter vacuum RBH.

\subsubsection{Relation with thin-shell gravastar}
The interesting feature of this particular form of EoS is that one can approach the original thin-shell gravastar model; that is, by using
\begin{equation}
    \bar{\epsilon}(y) = \bar{\epsilon}_0 = \frac{3}{8\pi} \frac{\alpha}{\bar{R}^3}\; (\text{constant}), \qquad 0\leq r\leq\bar{R}
\end{equation}
with a slightly different form of $\mathcal{T}(y)$ from Eq.~\eqref{eq. F(r) tanh epsilon}, that is
\begin{equation}
    \mathcal{T}(y) = \frac{1}{2}\left[1 + \tanh\left(\frac{y-\alpha\omega}{\alpha\sigma_t}\right)\right],
\end{equation}
and setting
\begin{equation}
    a = 1, \qquad \gamma = 1, \qquad \omega \to \frac{\bar{R}}{\alpha}, \qquad \sigma_t \to 0.
\end{equation}
With these parameter, the $F(y)$ behaves as a jump function,
\begin{equation}
    F(r) = 
    \begin{cases}
        0, & 0\leq y < \omega,\\
        2, & \omega \leq y \leq \bar{R}.
    \end{cases}
\end{equation}
Hence, it can be seen that the inner positive pressure boundary will be located at $\omega$ and the EoS at $\omega<y<\bar{R}$ becomes that of thin shell gravastar $p = \epsilon$. Since $F(\bar{R})=2$, which is the consequence of $\gamma=1$, we can directly conclude that the thin shell gravastar model violates the weak energy condition of transverse pressure at the surface.

It should be noted that this approach could only produce numerical thin shell gravastar solution where the thin-shell does not contribute to the total mass of the star. It comes from the fact that we should use numerical limit $\omega \to \bar{R}/x$ to make a thin shell, so mass contribution from the energy density in the thin shell is essentially zero. We can, however, relax the value of $\omega$ to a slightly lower value, but then it produces a thick shell gravastar with step transition rather than the thin shell one.

\section{Optical appearance}
\label{sec. IV optical appearance}
In this particular section, we use the line element in form of
\begin{equation}
    ds^2 = -A(r)dt^2 + B(r)dr^2 + r^2d\Omega.
    \label{eq. SSS AB}
\end{equation}

\subsection{Photon trajectories}\label{sec. photon trajectory}
We start from the geodesic equation,
\begin{equation}
    \frac{d\dot{x}^{\mu}}{d\tau} + \Gamma^{\mu}_{\alpha\beta} \frac{dx^{\alpha}}{d\tau} \frac{dx^{\beta}}{d\tau} = 0,
\end{equation}
where \(\tau\) is some affine parameter with \(x^{\mu}=(t,r,\theta,\phi)\) and \(\dot{x}^{\mu}\equiv dx^{\mu}/d\tau\). 
Solving the geodesic equation with spacetime in form of Eq.~\eqref{eq. SSS AB} on the equatorial plane (\(\theta = \pi/2,\, \dot{\theta} = 0\)), we obtain two conserved quantities
\begin{equation}
    A(r)\dot{t} = E, \qquad r^2\dot{\phi} = L.
    \label{eq. const of motion}
\end{equation}
With the 4-velocity condition of null geodesic, we eventually obtain the geodesic equation for a light ray
\begin{equation}
    \frac{d\phi}{dr} = \frac{1}{r^2} \sqrt{\frac{A(r)B(r)}{1/b^2 - V_p(r)}},\qquad V_p(r) \equiv \frac{A(r)}{r^2},
    \label{eq. dphidr}
\end{equation}
where \(V_p(r)\) is the photon's effective potential, and \(b\equiv L/E\) the impact parameter.

The effective potential is powerful to study the light behavior around an ultracompact object. If there exist local extrema ($\partial_r V_p(r) = 0$), the spacetime possesses photon sphere(s). Horizonless spacetime, or even regular spacetime in general, would have even number of photon spheres~\cite{Cunha:2017qtt,Cadoni:2022chn}; pairs of unstable and stable, and its stability is determined by whether $\partial_{rr} V_p(r)$ at the photon sphere location is negative (unstable) or positive (stable). One exceptional case is when these unstable and stable photon sphere coincide, that is at a particular configuration where there exists a point where $\partial_r V_p(r) = \partial_{rr} V_p(r) = 0$, and it is called the marginally-stable photon sphere. 

In our case, the photon sphere configuration is strongly determined by $x$. There is a minimum value of $x$ for the spacetime to produce a photon sphere, such that at $x=x_m$, the spacetime produces a marginally-stable photon sphere. Within the parameter studied in previous section with $\bar{R} = 1.1$ and $\bar{R} = 1.5$, we found that $x_m \approx 0.65368$ and $x_m \approx 0.65635$, respectively. Additionally, for the photon sphere to be right at the surface, one have $x=x_{ps}\equiv2\bar{R}/3\alpha_c$. 

\begin{figure}[htbp!]
    \centering
    \includegraphics[width=0.49\textwidth]{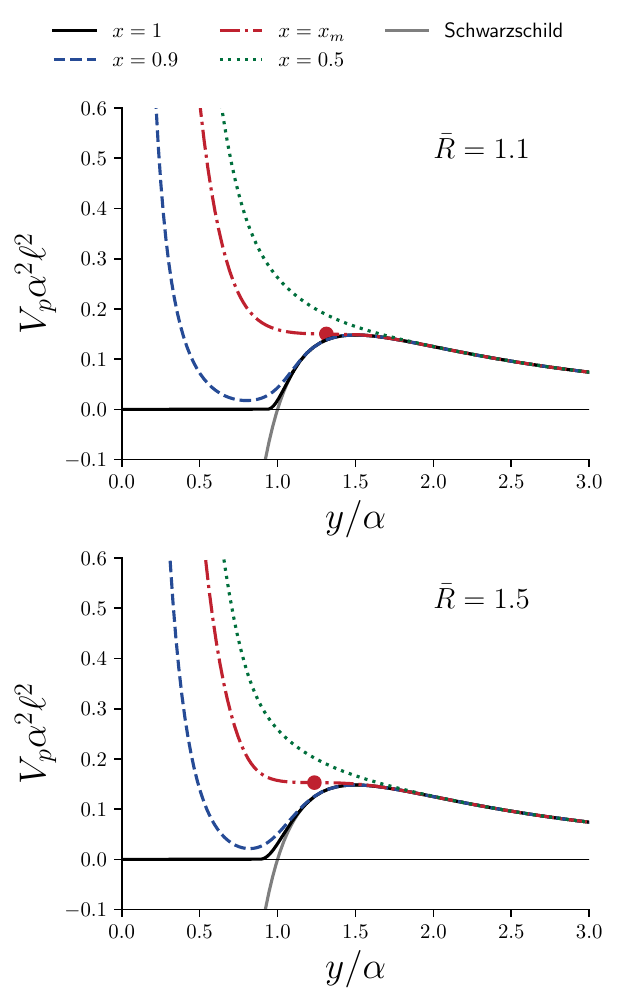}
    \caption{Effective photon potential for several values of $x$ with $\bar{R} = 1.1$ (top) and $\bar{R}=1.5$ (bottom). The red dot indicates the location of the marginally stable photon orbits with $x=x_m$.}
    \label{fig. effective photon potential}
\end{figure}

We plot our effective potential analysis in Fig.~\ref{fig. effective photon potential}. It can be seen that for both $\bar{R} = 1.1$ and $\bar{R} = 1.5$ with $x \geq x_{ps}$, the unstable photon sphere maintains its location at $y = 1.5\alpha$, or equivalently, $r = 3\mathcal{M}$, which is precisely the photon sphere radius of a Schwarzschild BH. This occurs because our star is ultracompact with $R < 3\mathcal{M}$, and since the exterior is a Schwarzschild vacuum, it results in a Schwarzschild photon sphere forming outside the star.

\begin{figure*}[htbp!]
    \centering
    \includegraphics[width=1\linewidth]{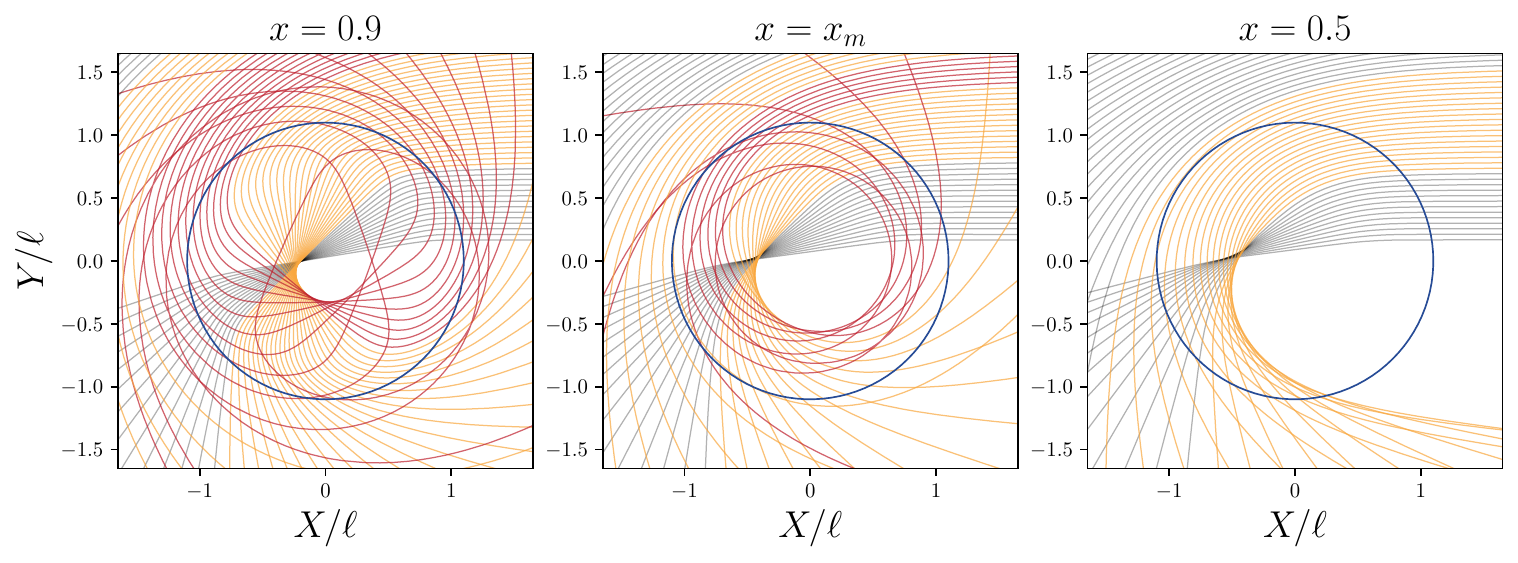}
    \caption{Photon trajectory around the object for several values of $x$ with $\bar{R}=1.1$. The direct ($n_r \leq 3/4$), lensed ($3/4 < n_r \leq 5/4$), and light ring ($n_r > 5/4$) trajectories are presented by the black, yellow, and red lines, respectively. The blue circle indicates the surface radius $\bar{R}$.}
    \label{fig. photon trajectory}
\end{figure*}

To see the photon behavior near the object, we plot the photon trajectory around the object in Fig.~\ref{fig. photon trajectory}. We classify three categories of photon trajectory (see, e.g., Refs.~\cite{Fauzi:2024nta,Fauzi:2024jfx,Zeng:2023fqy,Meng:2023htc}); The direct emission ($n_r \leq 3/4$), lensed emission ($3/4 < n_r \leq 5/4$), and light ring ($n_r > 5/4$),  where $n_r$ is the number of orbits described as
\begin{equation}
    n_r = \frac{\phi_{end}}{2\pi},
\end{equation}
with $\phi_{end}$ is the final angular position of the photon at the stopping point\textemdash which, in this case, is at numerical infinity.

In a configuration where at least one photon sphere exists, photon can `orbit' more than $5/4$ times around the spacetime before they sent back to infinity, shown by the red lines in Fig.~\ref{fig. photon trajectory}. In the absence of photon sphere, photon would only get lensed, but it might be possible to obtain a light ring trajectory when the configuration is near the formation of marginally stable photon sphere.

\begin{figure}[htbp!]
    \centering
    \includegraphics[width=0.45\textwidth]{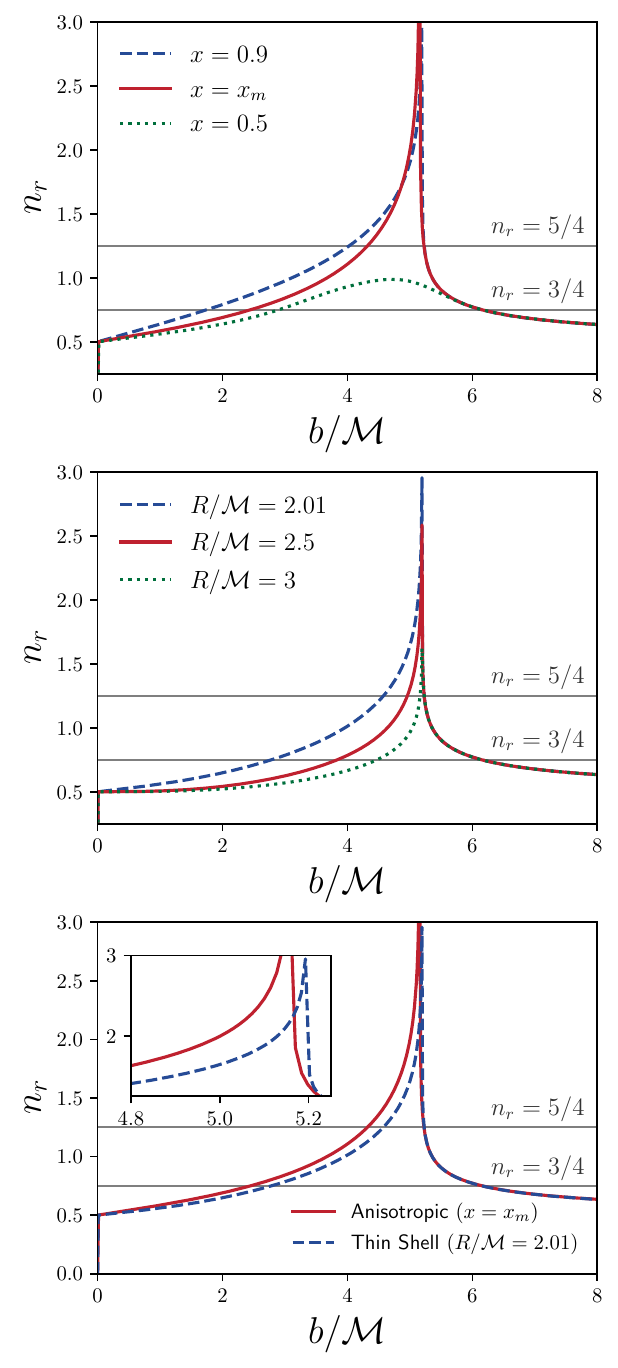}
    \caption{Number of photon orbits $n_r$ for a range of impact parameter: anisotropic gravastar with $\bar{R}=1.1$ (top), thin shell gravastar (middle), and comparison between anisotropic and thin shell gravastar with widest light ring impact parameter based on our model (bottom).}
    \label{fig. number of orbits}
\end{figure}
An interesting phenomenon also arise inside the object, which also known to appear in several other ultracompact object models; straight photon paths. This phenomenon occur in thin shell gravastar, where the photon paths is straight and slightly `lensed' once it reaches the surface. In semi-classical horizonless compact objects described by Ref.~\cite{Chen:2024ibc}, straight photon paths occur with smooth bending transition inside the object's surface, while in anisotropic gravastar described by Ref.~\cite{Fauzi:2024nta}, straight photon paths occur with smooth transition near the crust of the object. In our case, photon trajectories also transition into a straight path near the outer crust of the star. The reason on why straight path are produced in our gravastar model is fundamentally the same with the thin shell gravastar.

Consider the metric of thin shell gravastar in form of $e^{\nu(r)}=ke^{-\lambda(r)}$, where $k$ is related to the mass ratio of the thin shell mass and the volumetric mass, and it is known that $e^{-\lambda(r)}=1-2m(r)/r$ with $m(r) = 4\pi\epsilon_0 r^3/3$ (constant $\epsilon(r)=\epsilon_0$) \cite{Rosa:2024bqv}. The geodesic equation of Eq.~\eqref{eq. dphidr} now writes
\begin{equation}
    \phi = \int \frac{1}{r^2} \sqrt{\frac{1}{C-1/r^2}} dr,
\end{equation}
where $C\equiv(1/kb^2) + (8\pi\epsilon/3)$, and it can be integrated analytically, resulting in 
\begin{equation}
    r\cos(\phi-\chi)=\frac{1}{\sqrt{C}},
    \label{eq. straight path thin shell}
\end{equation}
 with $\chi$ an integration constant. This is just the equation of a straight line in polar coordinates, and the constant $C$ determines the closest distance of the line to the center. In our anisotropic gravastar, the profiles behave the same way; near the core region, the $e^{\Phi}$ component behaves as a constant, say $k$, less than unity as shown in Fig.~\ref{fig. metric component tanh}. We can also take the limit near the core for the mass function of Eq.~\eqref{eq. modified mass and energy density y}, yielding to $\lim_{y\to0}\bar{m}(y) = \bar{\kappa}\alpha y^3/2$. Hence, the equation of motion reads
 \begin{align}
     \phi &= \int \frac{1}{y^2}\sqrt{\frac{1}{1/(kb^2) - 1/y^2 + 2\bar{m}(y)/y^3}} dy & \notag\\
     &\approx \int \frac{1}{y^2}\sqrt{\frac{1}{C - 1/y^2}}dr, \quad C = 1/kb^2 + \bar{\kappa}\alpha& (y\to0).
     \label{eq. straight path anisotropic}
 \end{align}
However, this is also the case for original Hayward metric, where $e^{\phi}=k=1$. But, as shown in Ref.~\cite{Fauzi:2024nta} (and Ref.~\cite{Fauzi:2024jfx} for the modified Hayward metric), \textit{why do not we see the straight paths}? If one look closely into the photon trajectories with low impact parameter, the trajectories are \textit{almost} straight lines. Since the maximum value of $2\bar{m}(y)/y^3$ with original Hayward profile is $\bar{\kappa}\alpha$, the $1/kb^2$ term dominates in Eq.~\eqref{eq. straight path anisotropic} for photon with low impact parameters, so that at one point we can neglect the $2\bar{m}(y)/y^3$ term and eventually leads to an equation similar to Eq.~\eqref{eq. straight path thin shell}. In our anisotropic gravastar, the domination of $1/kb^2$ is more significant since $0<k<1$, hence the straight photon paths can be produced in slightly higher impact parameters.

\subsection{Thin accretion disk}
We assume an optically and geometrically thin accretion disk with monochromatic emission. For simplicity, we consider the accretion disk emission profile $I_e(r)$, which is related to the matter density and the temperature of the accretion disk. We adopt the emission profile known as the Gralla-Lupsasca-Marrone (GLM) model \cite{Gralla:2020srx},
\begin{equation}
    I_e(r) = \frac{\exp\left\{-\frac{1}{2}\left[\xi + \operatorname{arcsinh}{\left(\frac{r-\mu}{\sigma_j}\right)}\right]^2\right\}}{\sqrt{\left(r-\mu\right)^2+\sigma_j^2}},
    \label{eq. intensity profile glm}
\end{equation}
where $\xi$, $\mu$, and $\sigma_j$ are some parameters that determines the shape of the accretion disk. In contrast to exponential cut-off models (e.g., Refs.~\cite{Meng:2023htc,Zeng:2023fqy,Guo:2022iiy}), the GLM profile provides a continuous intensity function with adjustable smoothness, which is governed by the parameters $\sigma_j$ and $\xi$. This model has been shown to closely align with observational predictions of intensity profiles for astrophysical accretion disks, as derived from general relativistic magnetohydrodynamics~\cite{Vincent:2022fwj}, and has been applied in various studies of BHs and ultracompact objects' appearances~\cite{Rosa:2023hfm, Rosa:2024bqv, daSilva:2023jxa,Fauzi:2024nta}.

In this study, we consider two types of emission profiles:
\begin{enumerate}
    \item \textbf{GLM1}; characterized by $\xi = -2$, $\mu = R_{ISCO}$, and $\sigma_j = \mathcal{M}/4$, with $R_{ISCO}$ is the radius of the innermost stable circular orbit (ISCO) for a massive particle. This type of accretion disk assumes that the emission peaks and stops near the ISCO radius. This model is considered because any massive particle will eventually fall towards the center if it passes within the ISCO radius, leading to peak matter density around the ISCO radius. For general form of static and spherically symmetric spacetime, the \(R_{ISCO}\) can be calculated numerically by solving~\cite{Gao:2023mjb}
    \begin{equation}
    \left.3A(r)A'(r) - 2rA'(r)^2 + rA(r)A''(r)\right|_{r=R_{ISCO}} = 0.
    \label{eq. R Isco determine}
    \end{equation}
However, the configurations investigated in this study always satisfy $R < 6\mathcal{M}$. As a consequence, since the exterior spacetime is Schwarzschild, the innermost stable circular orbit (ISCO) remains the same as in the Schwarzschild case, namely $R_{\text{ISCO}} = 6\mathcal{M}$. Based on this, we can reasonably assume that the accretion disk profile near the ISCO should exhibit the same characteristics as that of a Schwarzschild BH.
	
    \item \textbf{GLM2}; characterized by $\xi=\mu=0$ and $\sigma_j = 2\mathcal{M}$. For this emission profile, we assume that the accretion disk spans through the center of the object, as there is no restriction for matter in horizonless spacetime to reach the center. The emission peaks at the center, $r=0$. However, this type of accretion disk will not be considered for inclined observations. Spanning the accretion disk up until the center is problematic within our scope if one consider the accretion flow effect, as it diverges at the central point. Hence, we will only use the GLM2 emission profile on the axial observation, where the accretion flow is neglected (static accretion disk) and the redshift effects only come from the gravitational redshift.
\end{enumerate}

In horizonless configurations, it has been argued that both of these profiles may exhibit additive properties. Since the ISCO feature is present while the horizon is absent, both intensity profiles can coexist, potentially intersecting and adding to each other~\cite{Rosa:2024bqv}. A detailed analysis of the accretion disk profile will not be pursued further, as it lies beyond the scope of our discussion. A relevant studies on accretion disk physics around a horizonless compact objects, particularly boson stars, can be found in Ref.~\cite{Rosa:2024eva}.

We incorporate the motion of the accretion flow to obtain a more realistic image. The accretion flow affects the observed intensity of the disk due to the relativistic Doppler effect of incoming and outgoing matter relative to the observer. We assume Keplerian accretion flow motion on the equatorial plane, with its four-velocity expressed as
\begin{equation}
    u^{\mu}_e = (u^t,u^r,0,u^{\phi}).
    \label{eq. four velocity accretion}
\end{equation}
Outside the ISCO radius, we assume the accretion flow does not experience any radial drift, so $u^r=0$. With the spacetime metric in form of Eq.~\eqref{eq. SSS AB}, the $u^{\phi}$-component is given by~\cite{Ryan:1995wh,Ozel:2021ayr}
\begin{equation}
    u^{\phi}_e(r) = \frac{\sqrt{A'(r)}}{\sqrt{2A(r)r - A'(r)r^2}}.
\end{equation}
The $u^{t}$-component can be obtained by the four-velocity condition for timelike particles, so that
\begin{equation}
    u^{t}_e(r) = \frac{\sqrt{2}}{\sqrt{2A(r) - A'(r)r}}.
\end{equation}
The observed photon frequency $\nu_o$ is weighted to the emitted frequency $\nu_e$ by the energy correction factor $\tilde{g}$ written as \cite{Ozel:2021ayr,M:2022pex,Kumaran:2023brp}
\begin{equation}
\nu_o = \tilde{g} \nu_e, \qquad
    \tilde{g} = \frac{-k_{\nu}u^{\nu}_o}{-k_{\mu}u^{\mu}_e},
    \label{eq. energy corr f}
\end{equation}
where $k_{\mu} = g_{\mu\nu}k^{\nu}$ is the four-velocity of the light ray and $u^{\nu}_o$ is the four-velocity of the observer. Since we assume that the far away observer is at rest in a flat spacetime, we have $u^{\nu}_o = (1,0,0,0)$ so that
\begin{equation}
    \tilde{g} = \left(u^t_e - \frac{k_{\phi}}{E}u^\phi_e\right)^{-1}.
\end{equation}
If we consider a static accretion disk where $u^{\phi}_e = 0$, we will have $\tilde{g} = 1/u^t_e = \sqrt{A(r)}$. In this case, it only calculates the effects of gravitational redshift. We will impose this scenario for axial observation images.

The light intensity with certain frequency $I^{\nu}$ are related to the frequency through the Lorenz invariant as $I^\nu_e/\nu_e^3 = I^\nu_o/\nu_o^3$ \cite{M:2022pex}. Integrating the intensity through the whole frequency range, we obtain the total observed intensity
\begin{equation}
    I_o(r) = \int I^\nu_o(r) d\nu_o = \tilde{g}^4 I_e(r).
\end{equation}

\subsection{Ray tracing images}

\begin{figure}[htbp!]
    \centering
    \includegraphics[width=0.35\textwidth]{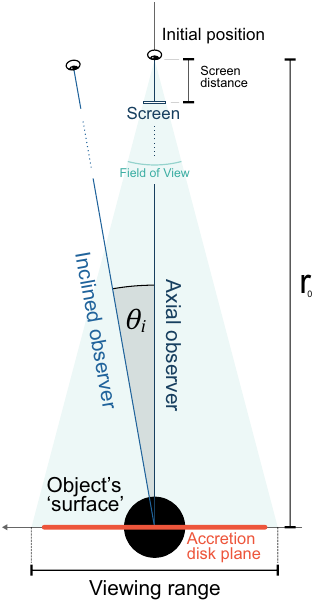}
    \caption{Two-dimensional illustration of axial and inclined observations, with $r_0$ is the observer distance and $\theta_i$ is the inclination angle.}
    \label{fig. ilustration inclination angle}
\end{figure}

Two approaches are employed for the ray tracing procedure in this study, as illustrated in Fig.~\ref{fig. ilustration inclination angle}: the \textit{axial observation}, where the observer line of sight is perpendicular to the accretion disk, and the \textit{inclined observation}, with more general viewing angle. In the former, one can take the advantage of the spherical symmetry. Instead of calculating every pixel on the screen, one can calculate the geodesic on the middle part of the screen, and `rotate' the resulting intensity to obtain the full image of the object's optical appearance. This way, the required null geodesic calculation is significantly reduced, and one can generate higher resolution images with cheap computation cost. Nevertheless, the inclined observation is useful to generate a more realistic shadow image model since it also calculates the Doppler redshift effect caused by the accretion flow.

In this discussion, we first generate the images of the thin shell gravastar to show that our construction approach could reproduce the thin shell gravastar images obtained in Ref~\cite{Rosa:2024bqv}. Then, we generate the images of our star model, where we use the same configuration as in Sec.~\ref{sec. photon trajectory}. Without loss of generality, we set $\ell=1$ so that $\mathcal{M} = \alpha/2$, $r=y$, and $R=\bar{R}$. The observer distance from the center object is $r_0=500\mathcal{M}$.

\subsubsection{Axial observation}
For the axial observation, we first generate and analyze the thin-shell gravastar model based on our pressure ansatz with GLM1 accretion disk profile. Next, we generate and examine the images of our proposed model with both GLM1 and GLM2 accretion disk profile, then finally compare the results with thin-shell gravastar.

\begin{figure}[htbp!]
    \centering
    \includegraphics[width=0.35\textwidth]{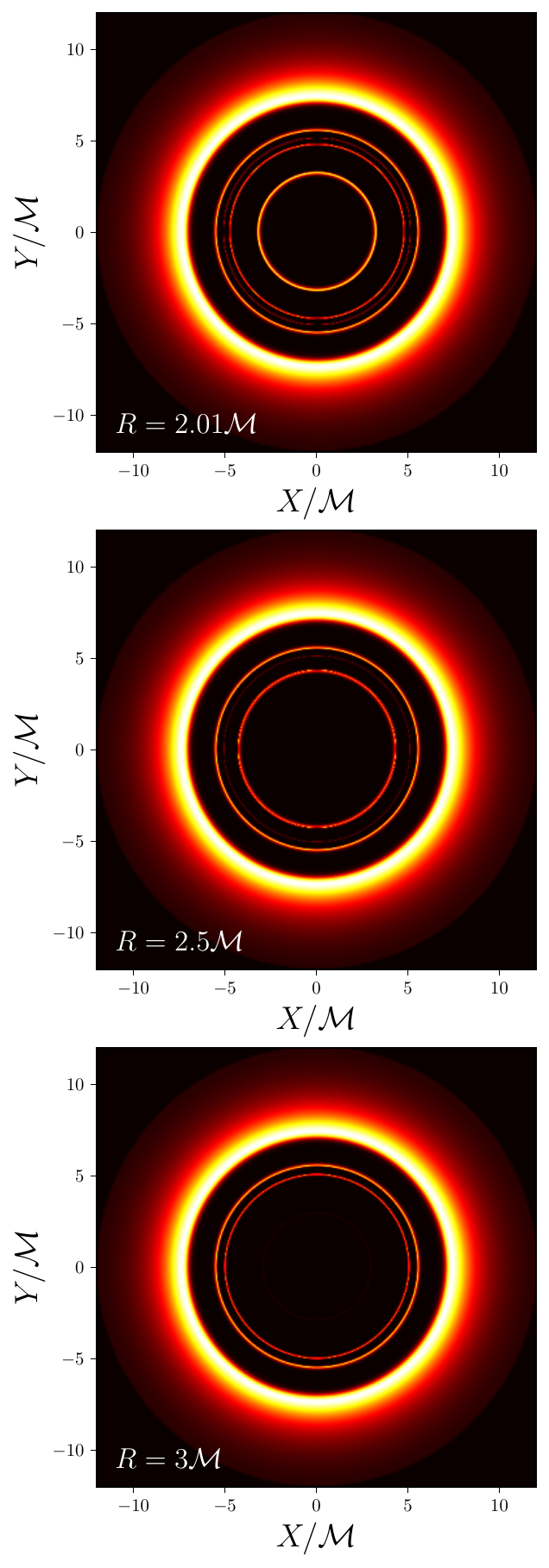}
    \caption{Optical appearence of thin shell gravastar based on our construction with axial observation and GLM1 accretion disk profile.}
    \label{fig. axial thin shell}
\end{figure}

\begin{figure*}[htbp!]
    \centering
    \includegraphics[width=0.8\textwidth]{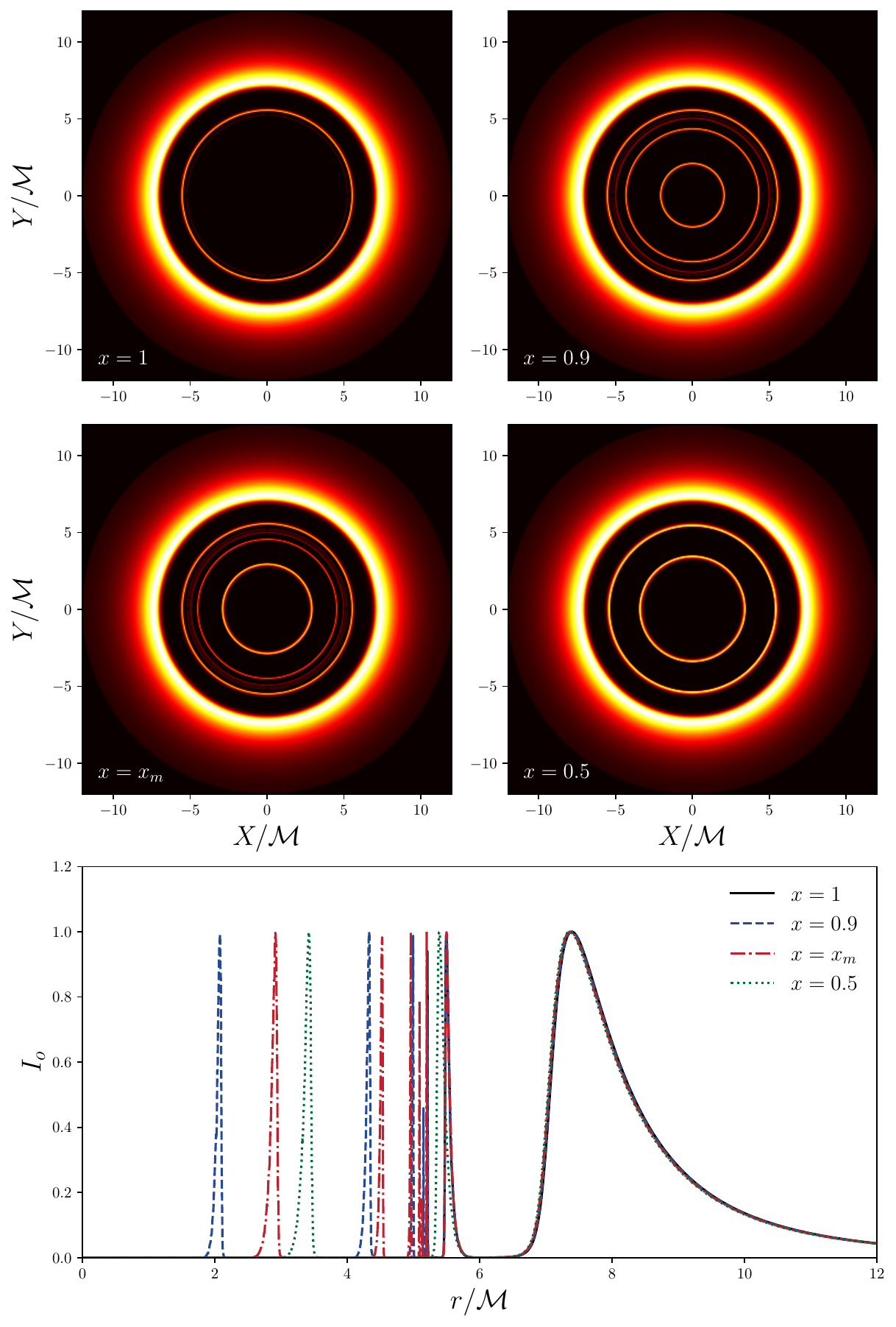}
    \caption{Optical appearence of our anisotropic gravastar model for axial observation with GLM1 (ISCO) accretion disk. The bottom plot represent the intensity cross-section for each images.}
    \label{fig. optical appearance axial}
\end{figure*}

\begin{figure*}[htbp!]
    \centering
    \includegraphics[width=0.8\textwidth]{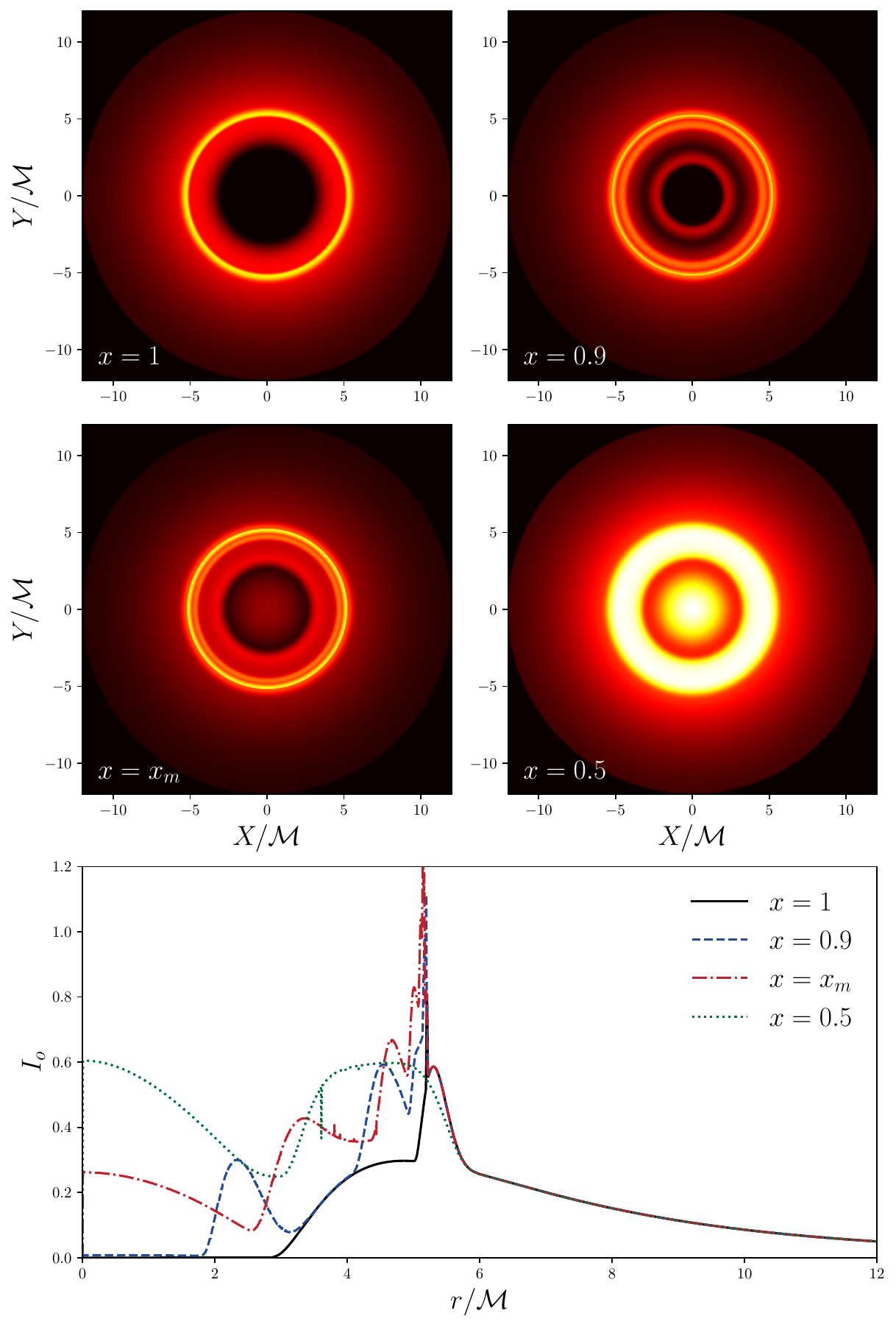}
    \caption{Optical appearence of our anisotropic gravastar model for axial observation with GLM2 (center) accretion disk. The bottom plot represent the intensity cross-section for each images.}
    \label{fig. optical appearance axial center}
\end{figure*}  
\textbf{GLM1}\textemdash We show the images of thin shell gravastar with $R\in\left\{2.01\mathcal{M},2.5\mathcal{M},3\mathcal{M}\right\}$ surrounded by accretion disk with GLM1 intensity profile in Fig.~\ref{fig. axial thin shell}. The thin shell model in this study does not contribute to the total mass of the gravastar, as previously mentioned. From the photon ring feature, it is similar to that of thin shell gravastar images produced by Ref.~\cite{Rosa:2024bqv}, where we observe three major photon rings on $R=2.01\mathcal{M}$ image with one thin photon ring between the first and second outer major photon rings, two major photon rings on $R=2.5\mathcal{M}$ and a thin photon ring between them, and also two major photon rings on $R=3\mathcal{M}$ with smaller seperation. The photon rings location, however, are slightly different from the previously mentioned reference. It should be tolerable since our approach employs strict numerical approximation.

We present our star model images along with their cross-section intensities in Fig.~\ref{fig. optical appearance axial}. The extremal configuration with $x = 1$ essentially shares the same optical appearance as the Schwarzschild BH, due to the presence of a Schwarzschild photon sphere outside the object (see Fig.~\ref{fig. effective photon potential} and the related discussion). Since the spacetime contains an (extremal) horizon, photons with an impact parameter smaller than that of the Schwarzschild photon sphere will eventually fall into the horizon. This results in a dark central region in the image, commonly referred to as the `shadow' feature (see Ref.~\cite{Fauzi:2024jfx} for a related discussion). 

In general, all horizonless configurations with $x<1$ share one common feature: an innermost photon ring, which becomes larger as the value of $x$ decreases. The presence of photon spheres also generates a noticeable signature\textemdash minor photon rings between the first and second outer photon rings \textemdash caused by the \textit{chaotic} behavior of photon trajectories near the light ring impact parameter. These minor photon rings are produced by light ring trajectories, as shown by the red lines in Fig.~\ref{fig. photon trajectory} and $n_r>5/4$ in Fig.~\ref{fig. number of orbits}, and are absent when $x=0.5$, where the spacetime does not contain any photon sphere.

Comparing the results from the thin-shell and our anisotropic gravastar, a question might arise: \textit{why do not we observe these chaotic minor photon rings in the thin shell gravastar images?} If one analyzes the photon trajectories inside a thin-shell gravastar from Ref.~\cite{Rosa:2024bqv}, it is essentially a straight line, and\textemdash compared to our anisotropic gravastar\textemdash only a small range of impact parameter close to the Schwarzschild critical impact parameter of $b_c=3\sqrt{3}\mathcal{M}$ produces a light ring trajectories (see Fig.~\ref{fig. number of orbits}). As a consequence, the only minor photon ring observed is the \textit{Schwarzschild ring}, which is high order secondary image produced by photon with Schwarzschild critical impact parameter.

\textbf{GLM2}\textemdash We also generate the images of our proposed star model with the GLM2 accretion disk intensity profile, shown in Fig.~\ref{fig. optical appearance axial center}. From these results, we can specifically analyze the role of gravitational redshift effects in the spacetime. In the extreme configuration with \( x = 1 \), we observe a dark patch at the center, indicating the apparent area of the horizon. In the horizonless configuration with \( x = 0.9 \), however, the dark patch is caused by the strong gravitational redshift near the center. As expected, the gravitational redshift effect at the center is not strong enough to produce a shadow feature in configurations with lower \( x \).

\subsubsection{Inclined observation}

\begin{figure}[htbp!]
    \centering
    \includegraphics[width=0.49\textwidth]{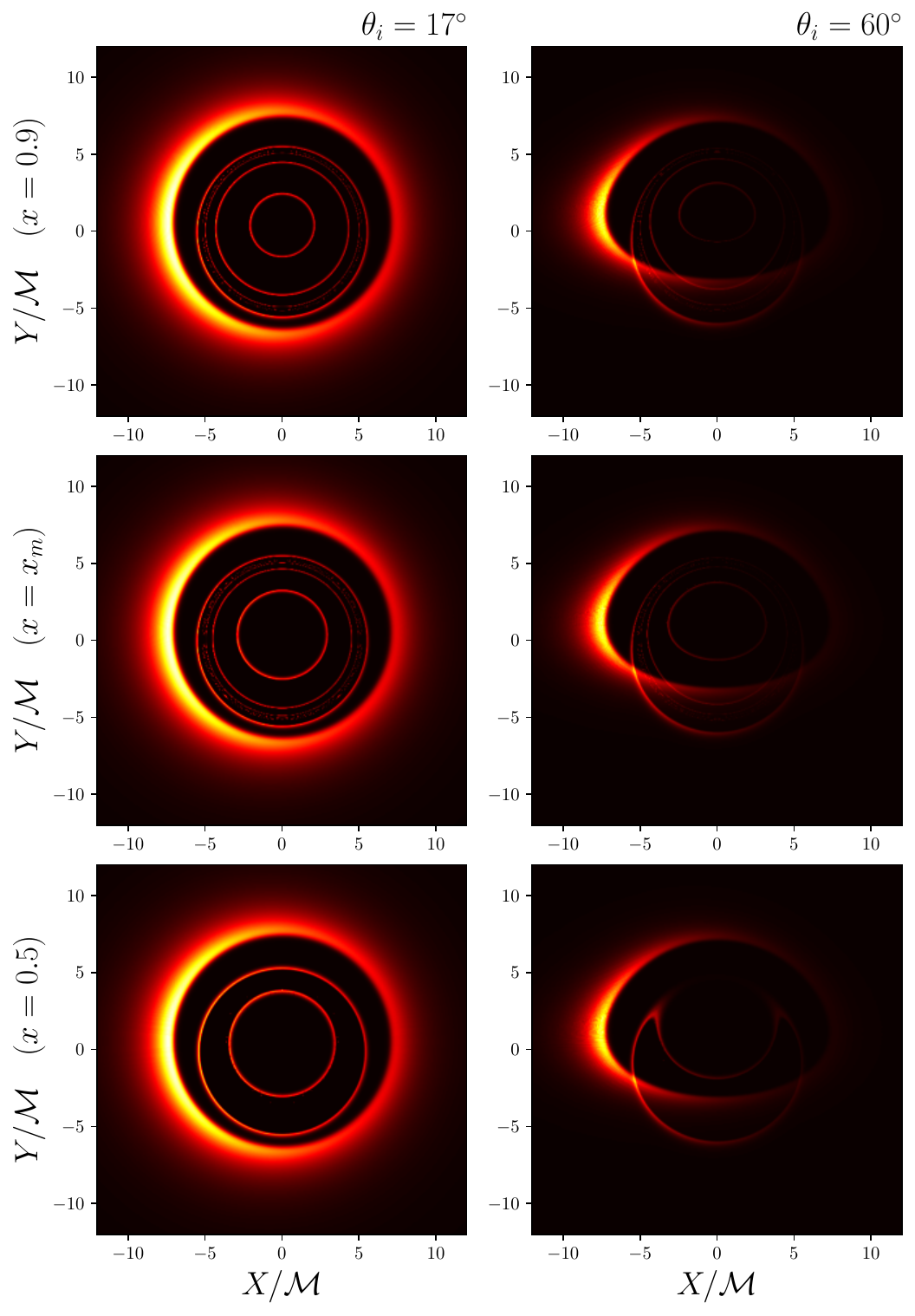}
    \caption{Optical appearence of our anisotropic gravastar model for inclined observation with inclination of $\theta_i=17^\circ$ (left) and $\theta_i=60^\circ$ (right), surrounded by GLM1 accretion disk profile.}
    \label{fig. optical appearance inclined}
\end{figure}

Here, we generate the images with inclined observation with inclination angle of $17^\circ$ and $60^\circ$. The accretion flow and redshift effects are taken into account to how it influence the appearence of the inner photon rings. We show our generated images in Fig.~\ref{fig. optical appearance inclined}. The photon rings still appears brightly at low inclination angle with a noticable brightness asymmetry between the right and the left side of the images caused by Doppler redshift effects. In higher inclination angle, the photon rings get dimmer normalized to its maximum intensity, especially with the higher $x$ value. One also observe a unique lensing phenomenon in the absence of photon sphere with $x=0.5$, where instead of two photon rings, it produce a crescent-like shape which is bright enough to be potentially observed.

These results indicate that observability of secondary images is in favorable with lower inclination angle, close to axial observation. The secondary images are also more visible in spacetime that does not feature any photon sphere, as long as the assumption that light does not interact with the object's interior still hold true. In a more realistic case, we might expect that the positive pressure region possibly emmits electromagnetic radiation which is potentially observed. However, the emitted radiation would be gravitationally-redshifted, and the observed intensity would be supressed at least by the factor of $\sqrt{A(r)}$. Thus, it is natural to assume that electromagnetic emission from the interior of the star is negligible, the especially with large $x$ configurations where the gravitational redshift effects is significant.

\section{Gravitational (axial) perturbation}
\label{sec. V gravitational perturb}

In this section, we study another phenomenology appearing in our horizonless star model. The effective potential and the quasinormal mode can be obtained from the perturbation method in Einstein gravity. In this case, we consider only the axial perturbations which lead us to the wave equation with the Reggae-Wheeler (RW) potential. For such perturbations, the very general metric can be written as\footnote{Important to note that this convention variable is only valid in this section.} \cite{Chandrasekhar:1985kt}
\begin{eqnarray}
    ds^2 &=& e^{2\nu}dt^2 - e^{2\psi} (d\varphi-\omega dt-q_{2}dx^2-q_{3}dx^3)^2 \nonumber \\ && -e^{2\mu_{2}}(dx^{2})^2 - e^{2\mu_{3}}(dx^{3})^2,
\end{eqnarray}
where $\nu$, $\psi$, $\mu_{2}$, and $\mu_{3}$ denote the unperturbed configurations while $\omega$, $q_{2}$, and $q_{3}$ denote the perturbation variable. The latter variables are functions of $t(=x^{0})$, $x^2(=r)$, and $x^3(=\theta)$. In the axial direction, the non-zero component of the Einstein field equation (in the tetrad frame) is [(1),(2)] and [(1),(3)]. The energy-momentum tensor with this component reads
\begin{eqnarray}
    T_{(1)(2)} &=& (\epsilon+p_{t})u_{(1)}u_{(2)}+p_{t}g_{(1)(2)}-(p_{t}-p)w_{(1)}w_{(2)}=0, \nonumber \\ T_{(1)(3)} &=& (\epsilon+p_{t})u_{(1)}u_{(3)}+p_{t}g_{(1)(3)}-(p_{t}-p)w_{(1)}w_{(3)}      =0,\nonumber \\ 
\end{eqnarray}
and the nonzero Ricci tensor can be written in a compact form 
\begin{eqnarray}
    -R_{(1)(2)}&=& \frac{1}{2}e^{-2\psi-\nu-\mu_{3}} \bigg[  (e^{3\psi+\nu-\mu_{2}-\mu_{3}} Q_{32}),_{3}  \nonumber \\ &&- 
 (e^{3\psi-\nu+\mu_{3}-\mu_{2}} Q_{02}),_{0}  \bigg]   , \\ \nonumber -R_{(1)(3)} &=&  \frac{1}{2}e^{-2\psi-\nu-\mu_{2}} \bigg[  (e^{3\psi+\nu-\mu_{3}-\mu_{2}} Q_{23}),_{2}  \nonumber \\ &&- 
 (e^{3\psi-\nu+\mu_{2}-\mu_{3}} Q_{03}),_{0}  \bigg],
\end{eqnarray}
where 
\begin{equation}
    Q_{AB}=q_{A,B}-q_{B,A}~~\textrm{and}~~ Q_{A0}=q_{A,0}-\omega_{,A}~~~(A,~B=2,~3).
\end{equation}
Hence, the perturbed Einstein field reads
\begin{eqnarray}
	(e^{3\psi-\nu-\mu_{2}-\mu_{3}}Q_{23}),_{3} + e^{3\psi-\nu-\mu_{2}+\mu_{3}}Q_{02,0} &=& 0, \nonumber \\ (e^{3\psi-\nu-\mu_{2}-\mu_{3}}Q_{23}),_{2} - e^{3\psi-\nu+\mu_{2}-\mu_{3}}Q_{03,0} &=& 0,
\end{eqnarray}
With the aid of these two latter equations, we can obtain
\begin{eqnarray}
	\label{req}
	&&[e^{-3\psi+\nu+\mu_{2}-\mu_{3}}(e^{3\psi+\nu-\mu_{2}-\mu_{3}}Q_{23}),_{3}],_{3} \nonumber \\ && + [e^{-3\psi+\nu+\mu_{2}+\mu_{3}}(e^{3\psi+\nu-\mu_{2}-\mu_{3}}Q_{23}),_{2}],_{2} = Q_{23,0,0}.
\end{eqnarray}
Eq. \eqref{req} can be separated by expanding the function in terms of Gegenbauer function $ \mathcal{C}_{n}^{\alpha}(\theta), $ that satisfying following differential equation
\begin{equation}
	\label{eq}
	\left[\frac{d}{d\theta}\sin^{2\alpha}\theta\frac{d}{d\theta}+n(n+2\alpha)\sin^{2\alpha}\theta\right]\mathcal{C}_{n}^{\alpha}(\theta)=0.
\end{equation}
By putting the separable equation defined as
\begin{eqnarray}
e^{3\psi+\nu-\mu_{2}-\mu_{3}}Q_{23}= X(r) ~\mathcal{C}_{l+2}^{-3/2}(\theta),
\end{eqnarray}
we find 
\begin{eqnarray}
    r^2 e^{\nu-\mu_{2}} \left( \frac{e^{\nu-\mu_{2}}}{r^2} X_{,r}\right)_{,r}-(l-1)(l+2) \frac{e^{2\nu}}{r^2}X+\sigma^2 X=0.
\end{eqnarray}
Further, by introducing the Tortoise radial coordinate
\begin{eqnarray}
    \frac{dr_{*}}{dr} = e^{\mu_{2}-\nu},~~~\textrm{and}~~~ X(r)=r\Psi(r),
\end{eqnarray}
Eq. \eqref{req} can be reduced to the time dependent wave equation in terms of $ r_{*} $
\begin{equation}
	\label{pdp}
	\left[\frac{\partial^2}{\partial t^2}-\frac{\partial^2}{\partial r_{*}^2}+V(r)\right]\Psi(r_{*},t)=0,
\end{equation}
where the effective radial potential is 
\begin{equation}
\label{veffint}
	V(y)\equiv V(r) \ell^2 = \frac{e^{2\nu}}{y^3}\left[l(l+1)y+4\pi y^3(\bar{\epsilon}-\bar{p})-6\bar{m}(r)\right]
\end{equation}
Note that we obtain the Regge-Wheeler equation with the pressure, density, and mass defined in a nonlocal manner. The second-order partial differential equation described above was endowed with three essential physical ingredients, i.e., the effective potential, complex eigenvalue, and the time dependence eigenfunction. Therefore, first, we want to analyze the effective potential behavior. Inside the star, we use the effective interior potential expressed in \eqref{veffint}, whereas, outside the star, we use the Schwarzschild exterior equation. 

\subsection{Effective potential}

\begin{figure}[htbp!]
    \centering
    \includegraphics[width=0.45\textwidth]{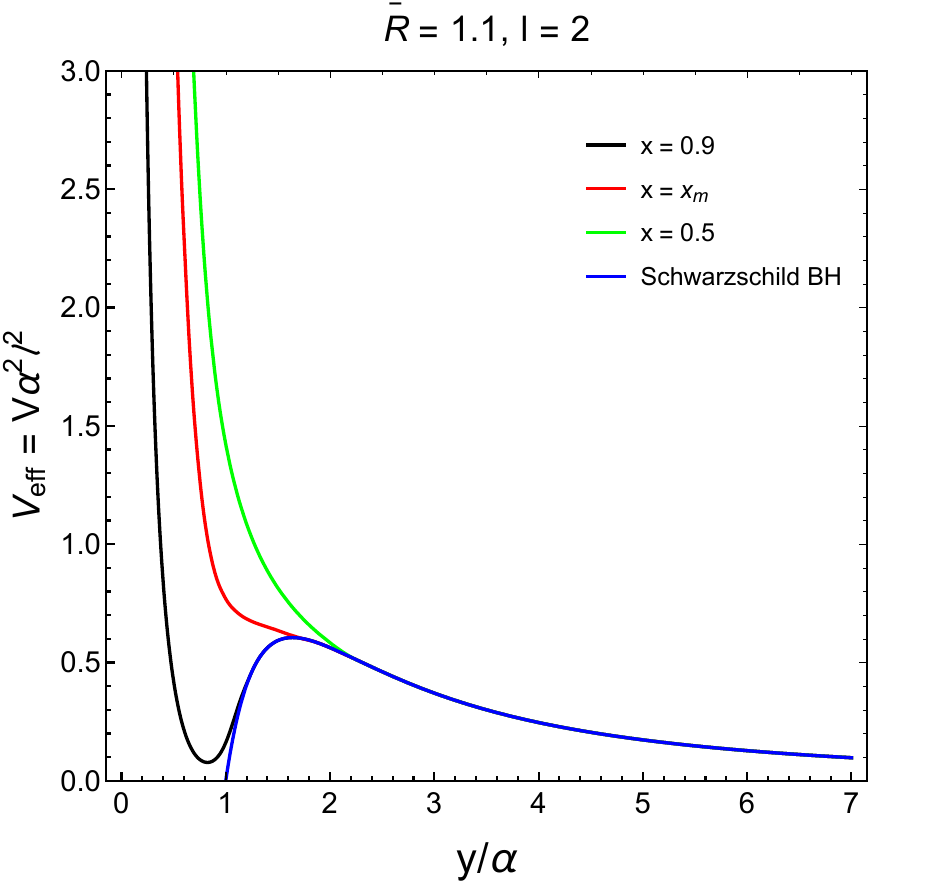}
    \includegraphics[width=0.45\textwidth]{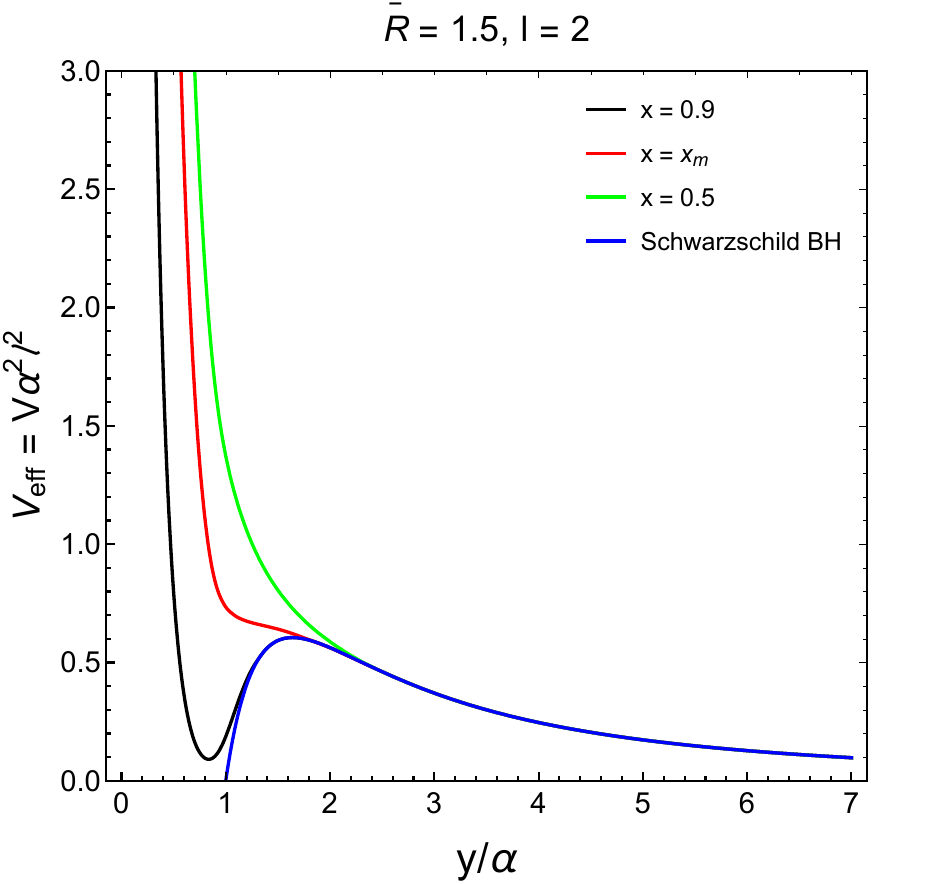}
    \caption{Axial GW effective potential profile of the anisotropic gravastar model with $l=2$. In BH case, the effective potential is $V 4\textrm{M}^{2}$ vs $r/2\textrm{M}$.}
    \label{fig:veff1}
\end{figure}

\begin{figure}[htbp!]
    \centering
    \includegraphics[width=0.45\textwidth]{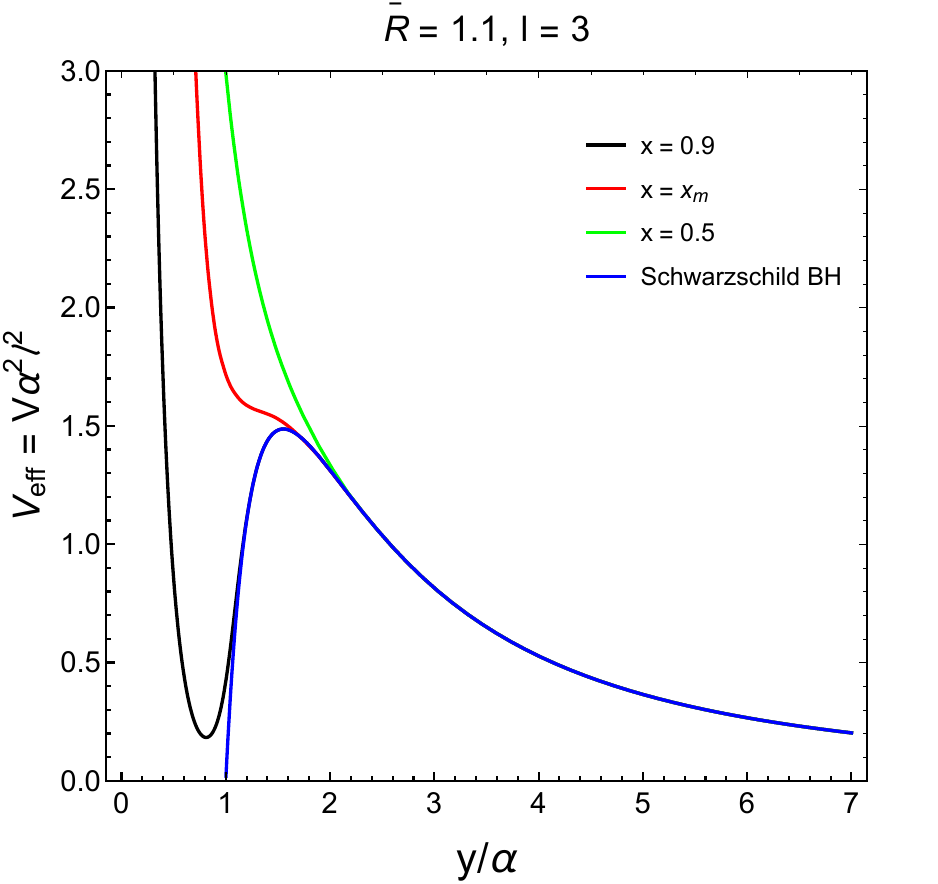}
    \includegraphics[width=0.45\textwidth]{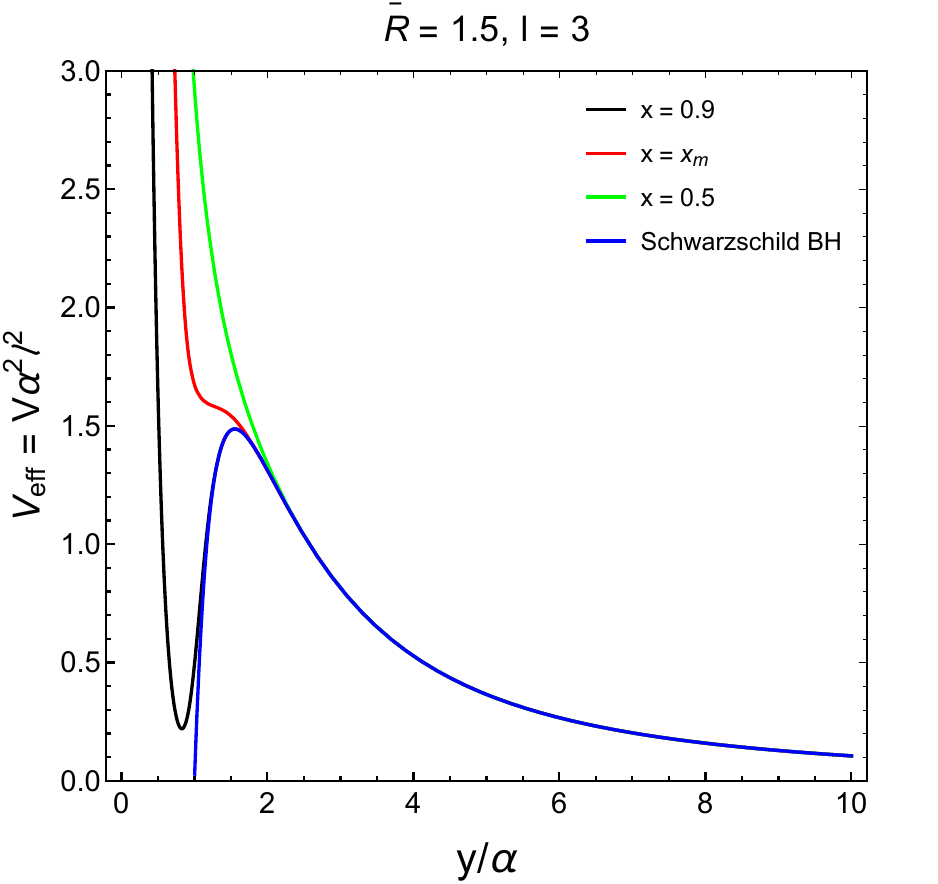}
    \caption{Axial GW effective potential profile of the anisotropic gravastar model with $l=3$. In BH case, the effective potential is $V 4\textrm{M}^{2}$ vs $r/2\textrm{M}$.}
    \label{fig:veff2}
\end{figure}

In Fig~\ref{fig:veff1} and Fig~\ref{fig:veff2}, we show the dimensionless effective potential as a function of $y/\alpha$. Here, we consider the effective potential of the horizonless star model with the smooth transition until the well fades away. In this figure, we also plot the effective potential for Schwarzschild BH case. The profile is matched with the anisotropic gravastar model. As we shift very close to the horizon $r/2M=1$, the potential drops, indicating the event horizon profile. The $x=x_{m}$ denotes the condition where the photon sphere starts to appear. In this condition, the potential well is not deep enough to trap quasinormal (trapped) mode even if we shift the $l$ values. On the other hand, a perturbed star's effective potential enables the infinite potential value to exist at the center due to the presence of the centrifugal contribution. The models with $x=0.9$ possesses a second photon sphere at the minimum of the effective potential. Since it has a positive value of the second derivative of the potential, the second photon sphere should be stable.

However, it is convenient to discuss the effective potential of the tortoise coordinate. The relation between the coordinate and the radius is
\begin{equation}
\label{ttt}
    \frac{dr_{*}}{dr}=\sqrt{-\frac{g_{rr}}{g_{tt}}},
\end{equation}
where $ g_{tt} $ and $ g_{rr} $ are the metrics in the interior and the exterior region of the star. We valuate Eq.(\ref{ttt}) numerically using the appropriate boundary condition and matching with the exterior tortoise coordinate.  

\begin{figure}[htbp!]
    \centering
    \includegraphics[width=0.45\textwidth]{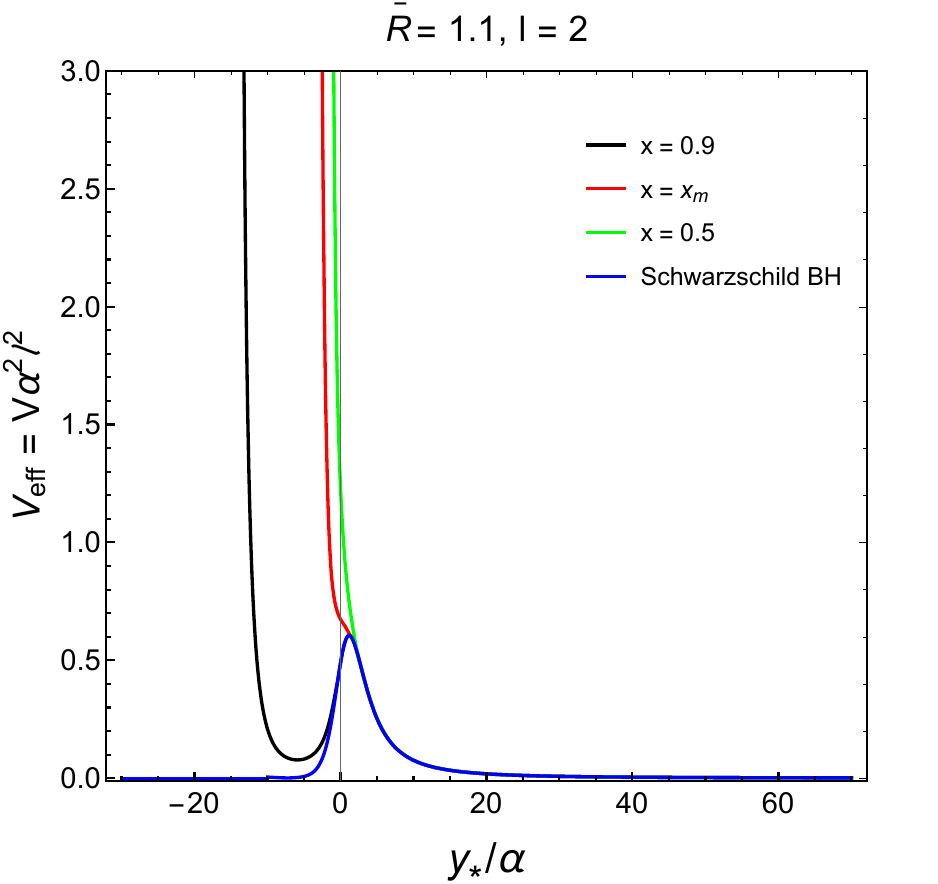}
    \includegraphics[width=0.45\textwidth]{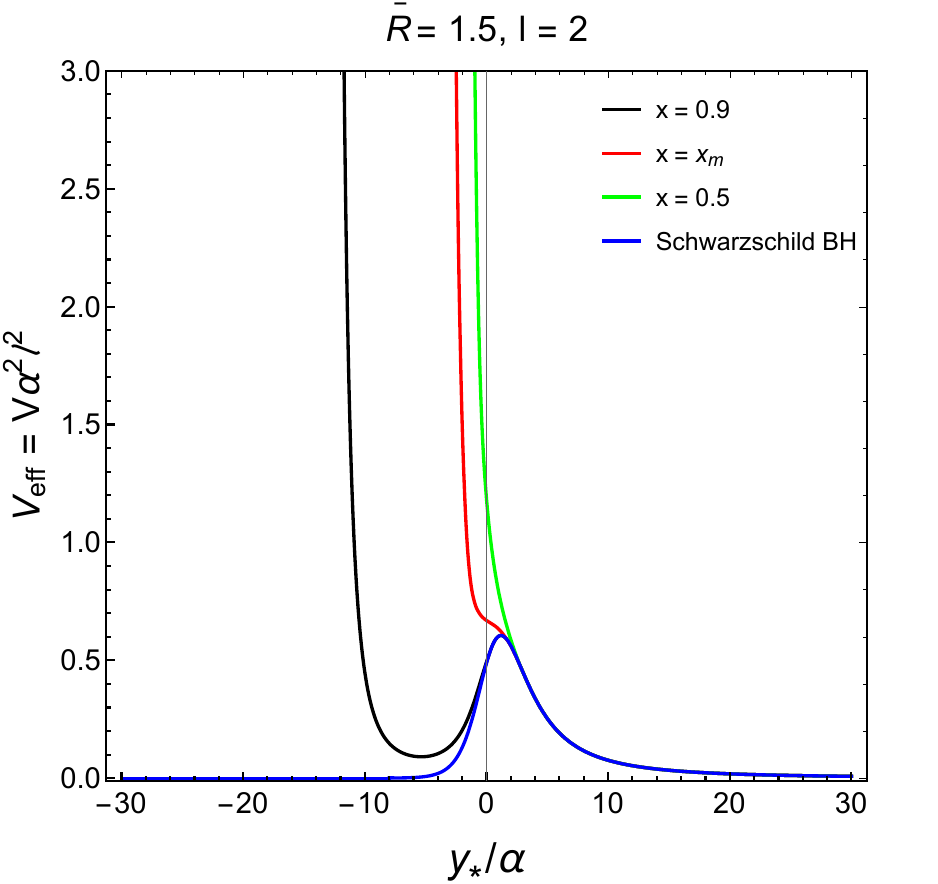}
    \caption{Effective potential profile vs tortoise coordinate $y_{*}$ of the anisotropic gravastar model with $l=2$. In BH case, the effective potential is $V 4\textrm{M}^{2}$ vs $r_{*}/2\textrm{M}$.}
    \label{fig:veffrstarl2}
\end{figure}

\begin{figure}[htbp!]
    \centering
    \includegraphics[width=0.45\textwidth]{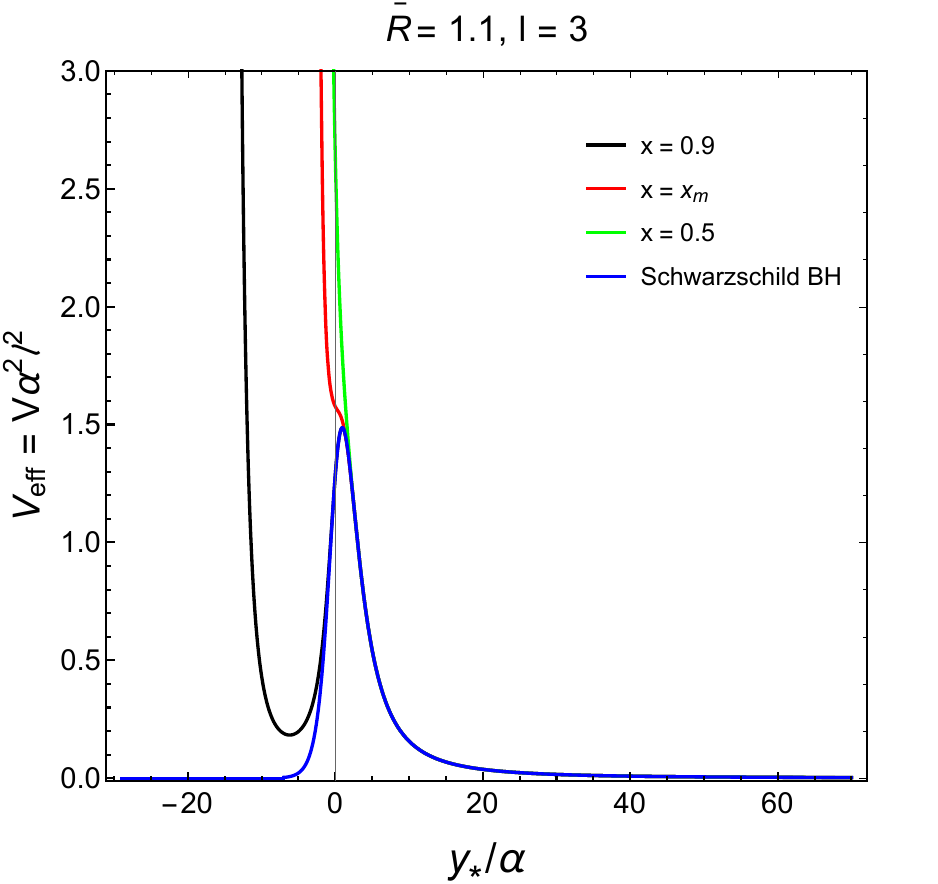}
    \includegraphics[width=0.45\textwidth]{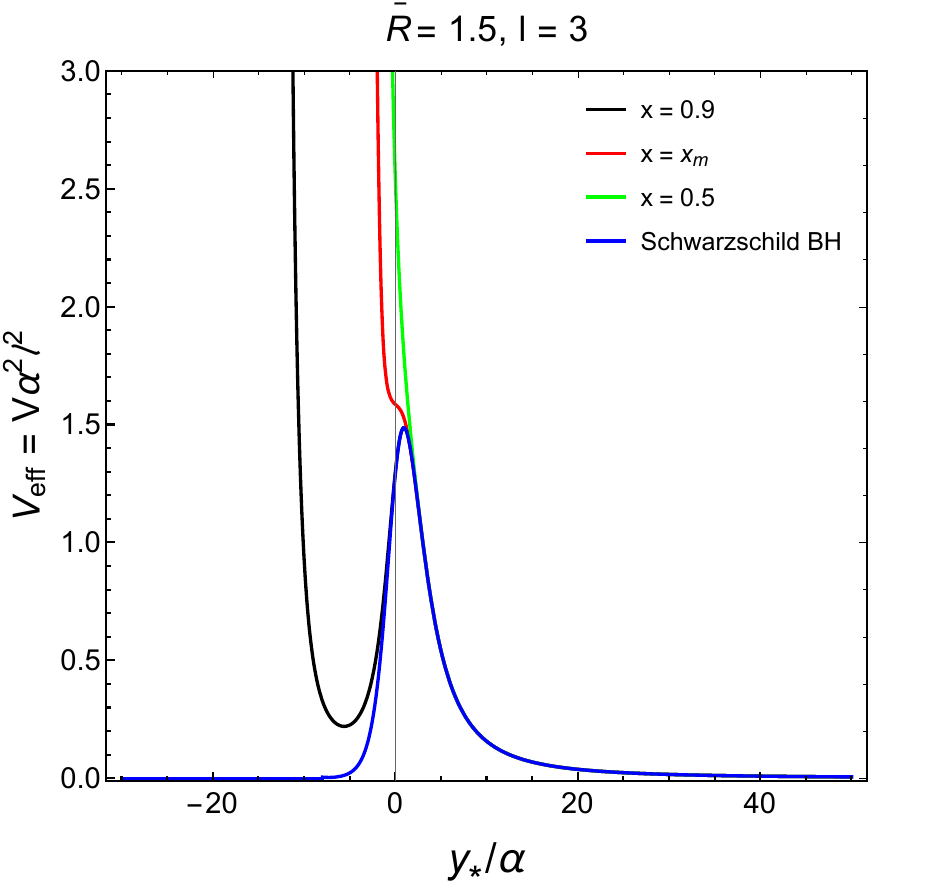}
    \caption{Effective potential profile vs tortoise coordinate $y_{*}$ of our anisotropic gravastar model with $l=3$. In BH case, the effective potential is $V 4\textrm{M}^{2}$ vs $r_{*
    }/2\textrm{M}$.}
    \label{fig:veffrstarl3}
\end{figure}

With the latter profile in hand, we can obtain the effective potential as a function of $ y_{*} $ plots. Fig.~\ref{fig:veffrstarl2} and Fig.~\ref{fig:veffrstarl3} show both interior and exterior regions of effective potential as a function of $ y_{*} $. One can see that the parameters $x$ and $\bar{R}$ on our model determines the width of the potential well. On the other hand, the photon sphere starts to exist when $x=x_{m}.$ Yet, the potential well is not deep enough to trap and resonate the incoming gravitational waves. When $x=0.5,$ the effective potential does not have a well at all.

These signatures, combined with several $x$ values, are important for the production of gravitational echoes as well as the quasinormal mode. Moreover, the earlier studies \cite{Kojima:1995nc,Andersson:1995ez} revealed that the spectrum mode related to the potential could contain three kinds of spacetime modes, i.e., the trapped modes \cite{Kokkotas:1994an} that correspond to the potential well predicted by horizonless stars, and the $ w $-modes \cite{Kokkotas:1992xak} that corresponds to the scattering on the top of the barrier, and the $ w_{II} $ mode \cite{Leins:1993zz} that corresponds to the extremely fast damped mode. In the next subsection, we will discuss only the mode crucial for our star model (axial part), i.e., the trapped mode.

\subsection{Quasinormal mode}

QNM, in general, is a dissipative property of spacetime and plays a crucial role in the form of gravitational wave echo when the stars reach the final stage during a ringdown phase. While the star spans these states, the unstable circular photon orbit traps the primary signal. In order to obtain the mode, the complex eigenvalue on the time-independent second differential equation must be solved with the appropriate boundary conditions. From the BH point of view (for example, Schwarzschild spacetime), since nothing can escape from them, we must employ an inward spherical boundary condition at the BH horizon
\begin{equation}
	\label{bc1}
	\Psi(r_{*},t)\approx e^{-i\omega(t+r_{*})},~~~~~~~r_{*}\rightarrow -\infty~(r\rightarrow 2M).
\end{equation}
We require a second boundary for the solution to be an outward wave at spatial infinity
\begin{equation}
	\label{bc2}
	\Psi(r_{*},t)\approx e^{-i\omega(t-r_{*})},~~~~~~~r_{*}\rightarrow \infty~(r\rightarrow \infty).
\end{equation}
The case is somewhat different from the ultracompact or compact stars' point of view since they lack the BH's event horizon (in this case, $\alpha<\alpha_{c}$). The first boundary condition in Eq. \eqref{bc1} should be replaced by the regularity condition at the center, whereas the second boundary condition in Eq. \eqref{bc2} is unaltered. Therefore, we can implicitly infer that the QNM spectrum for stars is quite different from the BH.

In this work, we follow the procedure in Ref.~\cite{Volkel:2017ofl} to obtain the QNM mode using the Wentzel-Kramers-Brillouin (WKB) approximation. In quantum mechanics, the Bohr-Sommerfeld (BS) rule is a well-known method to receive approximately for the energy spectrum, $ E_{n}, $ of bound states in a potential. With the WKB theory, it is probable to include higher-order correction of BS rule~\cite{popov}. It means that WKB approximation is the generalized version of the BS rule \cite{atomicwkb},
\begin{equation}
	\label{BS}
	\int_{x_{0}}^{x_{1}} \sqrt{E_{n}-V(x)} dx = \pi \left(n + \frac{1}{2}\right) - \frac{i}{4} e^{2i\int_{x_{1}}^{x_{2}} \sqrt{E_{n}-V(x)} ~dx},
\end{equation}
where $ x_{0} $ and $ x_{1} $ are the classical turning point(s) determined by the root of the integrand and also depend on the energy spectrum. The second term in Eq. \eqref{BS}  is the additional term described as the general one for the BS, where $ x_{2} $ is the third classical turning point on the right side of the potential barrier. For a detailed discussion of this procedure, please see Ref.~\cite{Volkel:2017ofl}. This additional term denotes an exponentially small imaginary part of the energy spectrum, which measures the barrier penetrability. We can simplify further by writing the energy spectrum into $ E_{n}=E_{0n}+iE_{1n} $, where $ E_{0n} $ is the real part of the energy whereas $ E_{1n} $ is small imaginary energy. Substituting these energy spectrums to the left-hand side of Eq.~\eqref{BS} and matching with the real and imaginary part at the right-hand side of Eq.~\eqref{BS}. The results are
\begin{equation}
	\label{e0n}
	\int_{x_{0}(E_{0n})}^{x_{1}(E_{0n})} \sqrt{E_{0n}-V(x)} dx = \pi \left(n + \frac{1}{2}\right),
\end{equation}
and
\begin{eqnarray}
	\label{e1n}
	E_{1n} &=& -\frac{1}{2}\left( \int_{x_{0}(E_{0n})}^{x_{1}(E_{0n})} \frac{1}{\sqrt{E_{0n}-V(x)}} dx \right)^{-1} \\ \nonumber
    && \times e^{2i \int_{x_{1}(E_{0n})}^{x_{2}(E_{0n})} \sqrt{E_{0n}-V(x)} ~ dx}.
\end{eqnarray}
Furthermore, we use the well-known analytic fitting potential described below \cite{Volkel:2017ofl}
\begin{eqnarray}
	U_{Q} = U_{0}+\lambda_{0}^2 (x-x_{min})^{4}, 
    \end{eqnarray}
    and
    \begin{eqnarray}
    U_{BW} = \frac{U_{1}}{1+\lambda_{1}(x-x_{max})^2}.\nonumber \\
\end{eqnarray} 
Both functions above are called the quartic oscillator and the Breit-Wigner potentials, respectively. These two functions depend on $ U_{0}, $ $ U_{1}, $ $ \lambda_{0}, $ and $ \lambda_{1}$ parameters. Then, the value of these parameters is matched with the true effective potential (in tortoise coordinate). The variable $ U_{0} $ will be the interior of the true potential, whereas the variable $ U_{1} $ is the one for the maximum of the true potential. Next, $ \lambda_{1} $ can be obtained by identifying $ V_{max}'' $ on the $ U_{BW} $ function, which can be written as $ \lambda_{1}=-V_{max}''/2V_{max}. $ The last parameter, $ \lambda_{0}, $ is demand that, the quartic oscillator must be equal to the Regge-Wheeler equation at the surface. However, we can start to evaluate the energy spectrum by inserting both fitting functions into Eq.\eqref{e0n} and Eq.\eqref{e1n}. The solution reads
\begin{eqnarray}
	\label{E0n}
	E_{0n} &=& U_{0} + \lambda_{0}^{2/3} \left[\frac{3\pi}{4~ \mathcal{K}(-1)} \left(n+\frac{1}{2}\right)\right]^{4/3}, \\ E_{1n} &=& -\frac{\sqrt{\lambda_{0}}(E_{0n}-U_{0})^{1/4}}{4~ \mathcal{K}(-1)} ~\exp\bigg[4i  \sqrt{\frac{U_{1}-E_{0n}}{\lambda_{1}}}~ \nonumber \\ && \mathcal{E}\left( i \sinh^{-1}\left(\sqrt{\frac{U_{1}}{E_{0n}}-1}\right),\frac{E_{0n}}{E_{0n}-U_{1}} \right)   \bigg],\nonumber \\
	\label{tuti}
\end{eqnarray}
where $ \mathcal{K}(a) $ is the complete elliptic integral of the first kind and $ \mathcal{E}(a,b) $ denotes the elliptic integral of the second kind \cite{Abraham}. The variable $ n $ shown in the Eq. \eqref{E0n} is the overtone number. We can match the energy with the complex eigenvalue which can be written as, $ \omega_{n}=\sqrt{E_{n}}=\sqrt{E_{0n}+iE_{1n}}=\omega_{r}+i\omega_{i}$, where the real part describes the frequency of the oscillation and the imaginary part describes the inverse of the damping time, $ \tau_{d} $. This brings us to the approximate analytic results for all trapped modes in our horizonless stars model.
\begin{figure}
    \centering
    \includegraphics[width=0.45\textwidth]{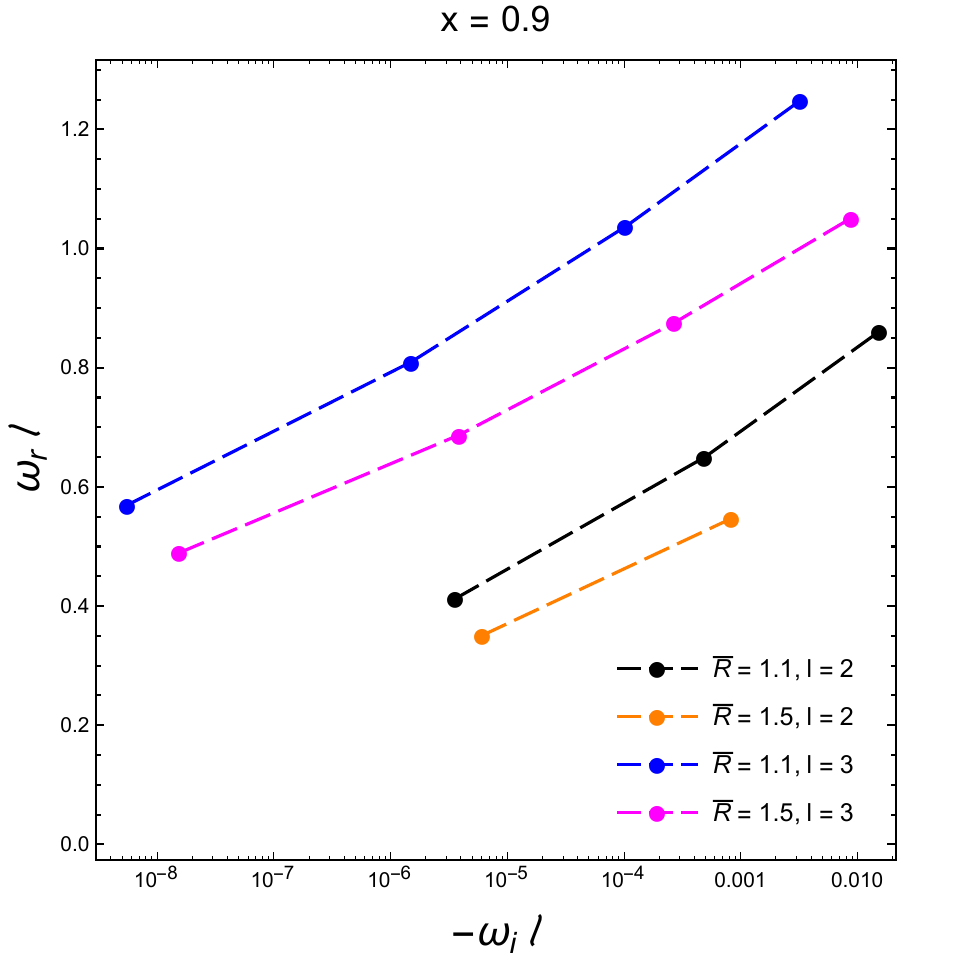}
    \caption{QNM profile for our anisotropic gravastar model with $x=0.9$}
    \label{fig:qnm}
\end{figure}

\begin{table}[h!]
\centering
\begin{tabular}{c c c c} 
 \hline\hline
 n ($l=2$) & $\omega\ell = \omega_{r}\ell+i\omega_{i}\ell$ & n ($l=3$) & $\omega\ell = \omega_{r}\ell+i\omega_{i}\ell$ \\ [0.5ex] 
 \hline
 0 & $0.413-3.68 \times 10^{-6}$ & 0 & $0.570-5.61\times 10^{-9}$ \\ 
 1 & $0.650-4.96\times 10^{-4}$ & 1 & $0.862-1.57\times 10^{-6}$ \\
 2 & $1.25-3.27\times 10^{-3} $ & 2 & $1.04-1.05\times 10^{-4}$ \\
 3 &  & 3 & $ 1.25-3.27\times 10^{-3} $\\ [1ex] 
 \hline
\end{tabular}
\begin{tabular}{c c c c} 
 \hline\hline
 n ($l=2$) & $\omega\ell = \omega_{r}\ell+i\omega_{i}\ell$ & n ($l=3$) & $\omega\ell = \omega_{r}\ell+i\omega_{i}\ell$ \\ [0.5ex] 
 \hline
 0 & $0.3514-6.23 \times 10^{-6}$ & 0 & $0.4901-1.59\times 10^{-8}$ \\ 
 1 & $0.5477-8.52\times 10^{-5}$ & 1 & $0.6874-3.91\times 10^{-6}$ \\
 2 &  & 2 & $0.8760-2.75\times 10^{-4}$ \\
 3 &  & 3 & $ 1.051-8.96\times 10^{-3} $\\ [1ex] 
 \hline
\end{tabular}
\caption{The numerical value of the quasinormal mode for our anisotropic gravastar model with $\bar{R}=1.1$ (top) and $\bar{R}=1.5$ (bottom).}
\label{table:1}
\end{table}

The result of the QNM is shown in Fig.~\ref{fig:qnm} and the exact values are presented in Table.~\ref{table:1}. We plot the QNM, $ |\omega_{i}| $ vs $ \omega_{r} $, with different $ l $. The number of the dots in the plot represents the overtone number. From these values, we can conclude that the profile with $ l=3 $ modes allows the ultracompact star to have more trapped modes. We can also infer that the horizonless ultracompact object with $\bar{R}=1.1$ allows the axial perturbations to damped longer than the star with $\bar{R}=1.5$. This is the consequence of the fact that the well of the star with $\bar{R}=1.1$ is wider.

\subsection{Gravitational echoes and echo time}
\begin{figure}[h!]
    \centering
    \includegraphics[width=0.45\textwidth]{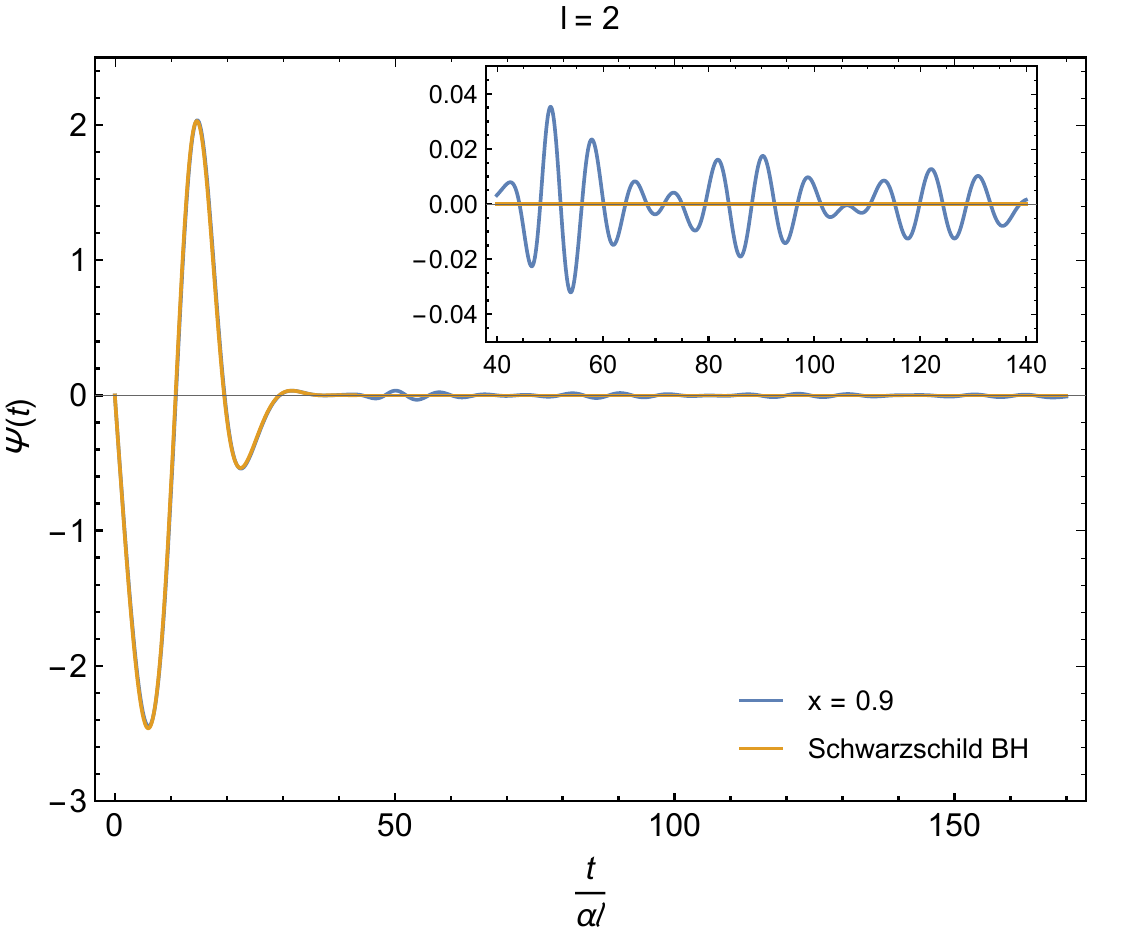}
    \includegraphics[width=0.45\textwidth]{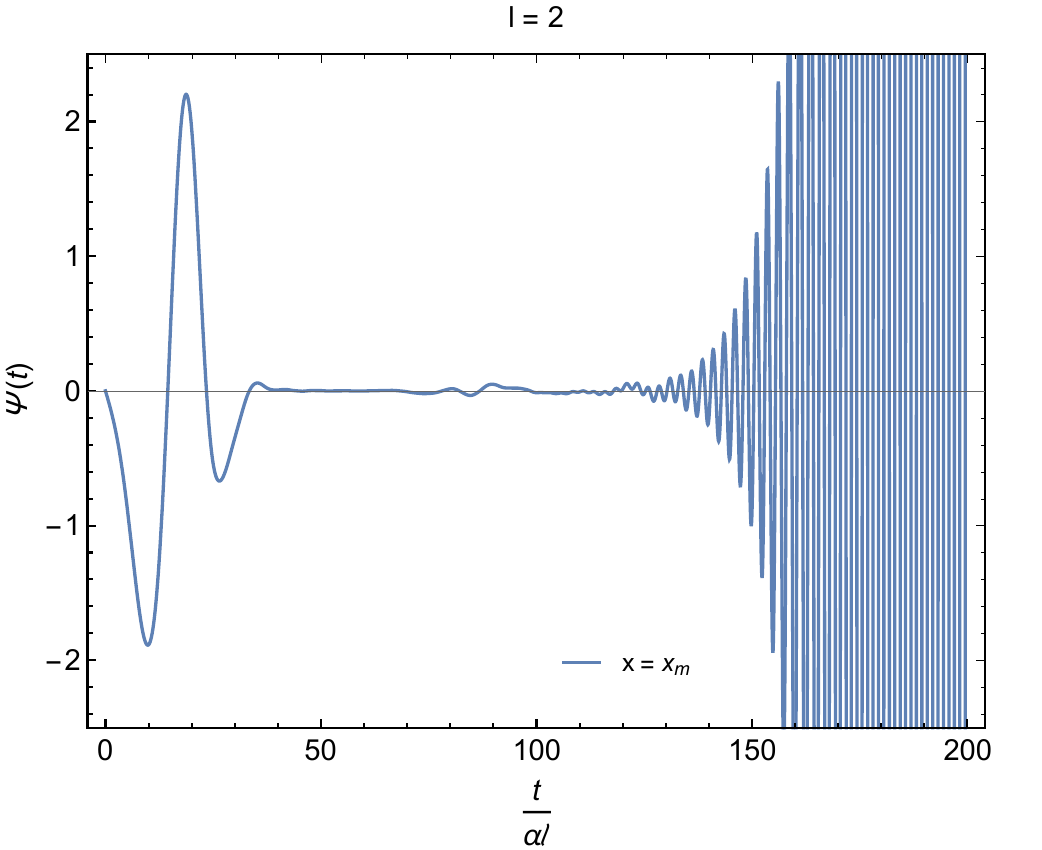}
    \includegraphics[width=0.45\textwidth]{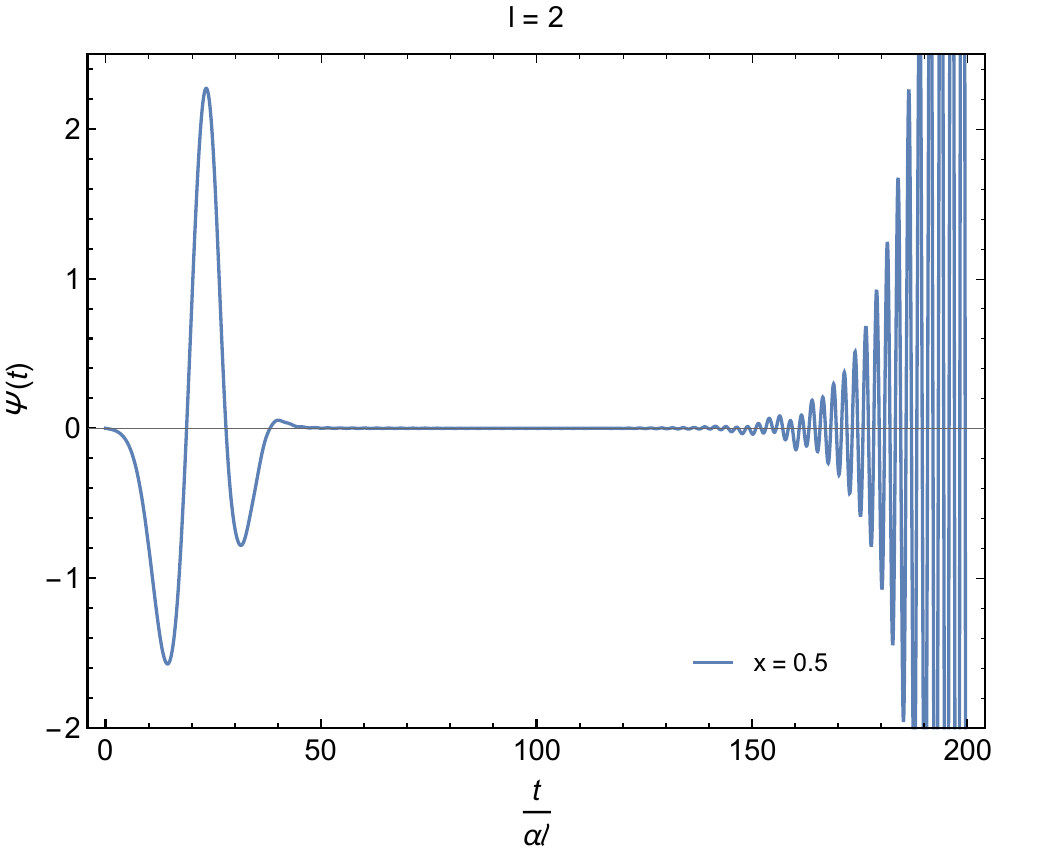}
    \caption{Gravitational echo profiles of the anisotropic gravastar model with $\bar{R}=1.1$. In BH case, the gravitational echoes is $\Psi(t)$ vs $t/2\textrm{M}$.}
    \label{fig:gravechoR11}
\end{figure}
\begin{figure}[h!]
    \centering
    \includegraphics[width=0.45\textwidth]{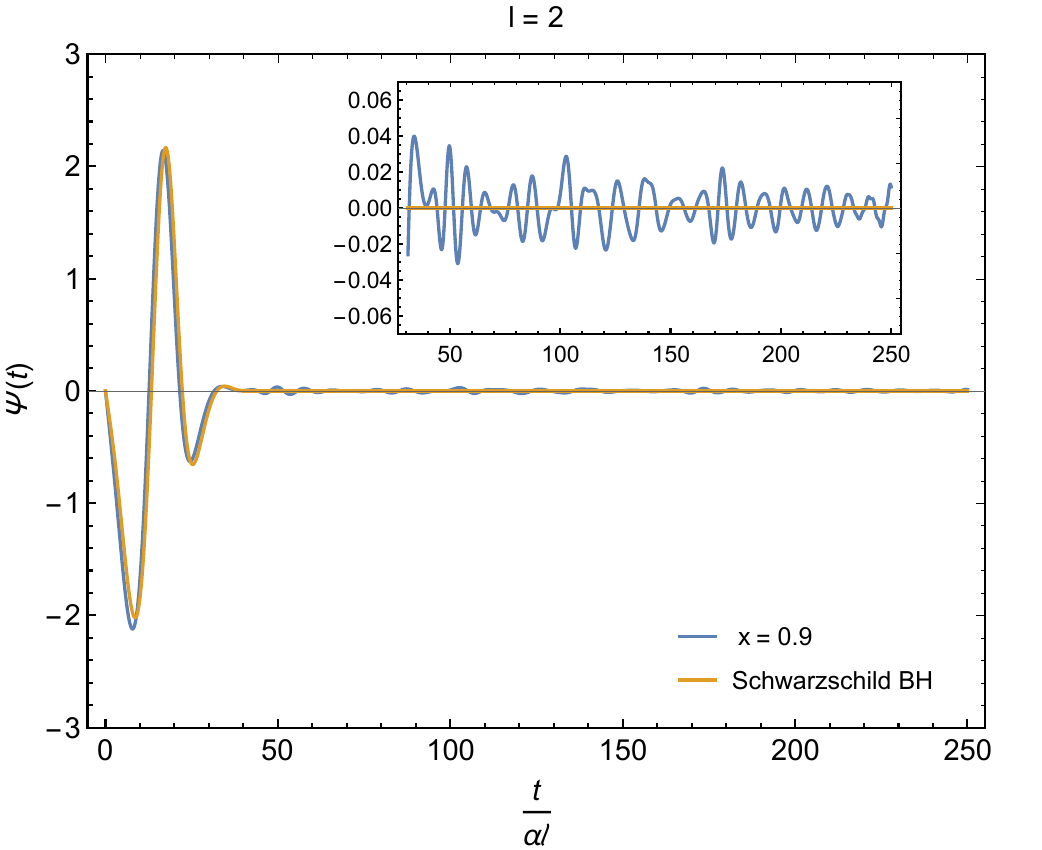}
    \includegraphics[width=0.45\textwidth]{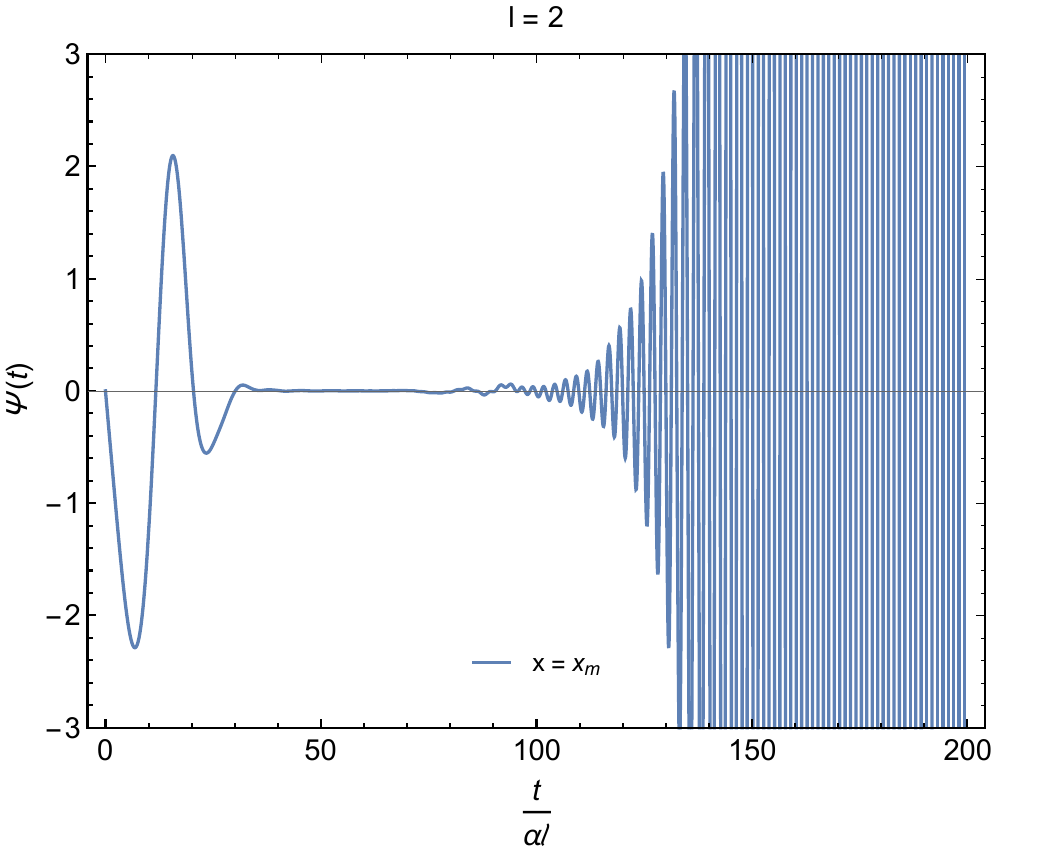}
    \includegraphics[width=0.45\textwidth]{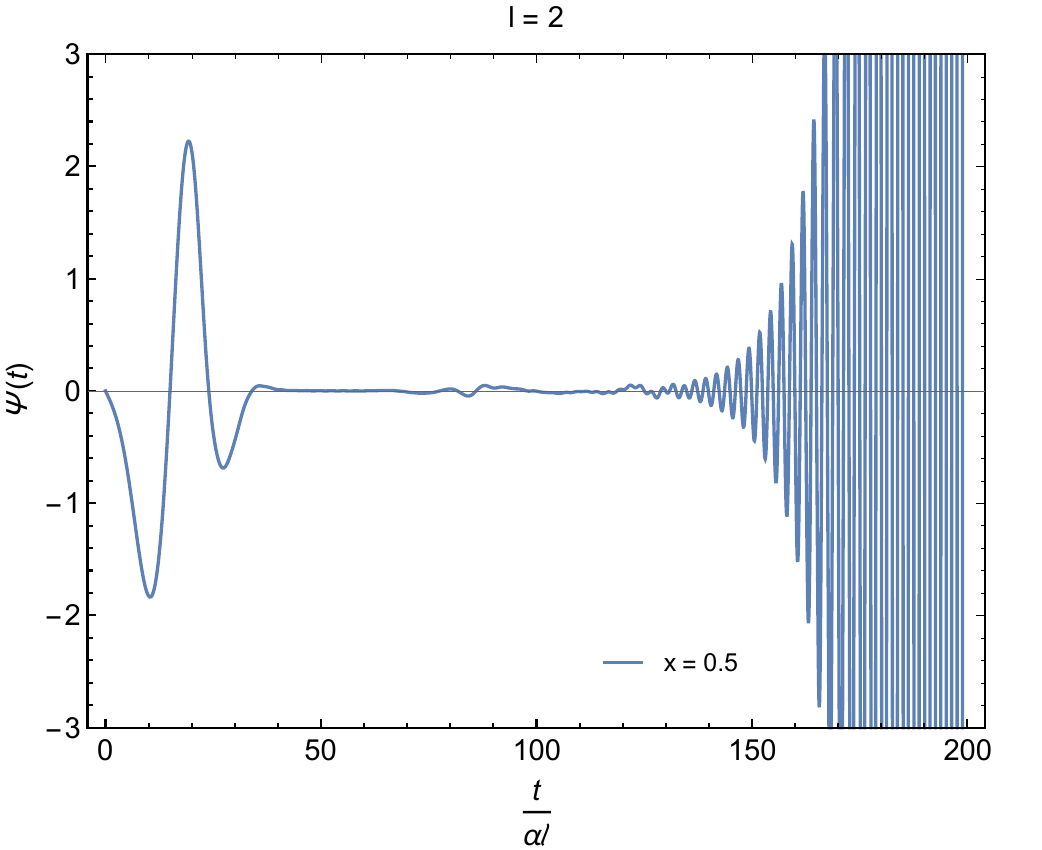}
    \caption{Gravitational echo profiles of our anisotropic gravastar model with $\bar{R}=1.5$. In BH case, the gravitational echoes is $\Psi(t)$ vs $t/2\textrm{M}$.}
    \label{fig:gravechoR15}
\end{figure}

In this section, we numerically analyze the gravitational echoes from the perturbed ultracompact star by solving the time-dependent partial differential equation in Eqn. \eqref{pdp}. Since the highest differential operator in the equation is of the second order in $t$ and $ r_{*}, $ the solution requires four conditions: 
\begin{eqnarray}
\textrm{I.}&&~\Psi(r_{*},0)~~~~~~~~~~=~0,\nonumber\\
\textrm{II.}&&~\Psi(r_{*}^{c},t)~~~~~~~~~~=~0,\nonumber\\ 
\textrm{III.}&&~\frac{\partial \Psi(t,r_{*})}{\partial t}\bigg|_{t=0}~~~=~f(r_{*}),\nonumber\\
\textrm{IV.}&&~\frac{\partial \Psi(r_{*},t)}{\partial r}\bigg|_{r_{*}\rightarrow \infty} = -\frac{\partial \Psi(r_{*},t)}{\partial t}\bigg|_{r_{*}\rightarrow \infty}.
\end{eqnarray}
Condition I and III are the initial data and II and IV are the boundary conditions. Both initial data denotes the post-merger phase with an initial Gaussian pulse centered at $ r_{*}=r_{g} $ and with spread $ \sigma_g $; $ f(r_{*})=\exp\big[-(r_{*}-r_{g})^2/\sigma_g^2\big] $ \cite{Urbano:2018nrs,Cardoso:2016oxy}. III. is the condition for the star to be regular at the center, where $ r_{*}^{c}(r)=r_{*}(0). $ Lastly, condition IV. relates outgoing waves at spatial infinity \cite{Urbano:2018nrs}.

After evaluating numerically, the solution of the differential equation depends on the time and tortoise radius, $ \Psi(y_{*},t). $ In this discussion, we focus on the time evolution of the signal and present only the lowest $(l=2)$ mode. The signal can be obtained by inserting the effective potential (as a function of tortoise coordinate $ y_{*} $) into the Eqn.~\eqref{pdp}. The results are presented in Fig.~\ref{fig:gravechoR11} and Fig.~\ref{fig:gravechoR15}. It is shown that the $ V_{\textrm{eff}} $ with $x=0.9$ in both $\bar{R}$ has a well and therefore it produces the gravitational echoes. The train wave after the merger state represents the signal that enters the interior and get reflected by the center of the star. On the other hand, the star with $x=x_{m}$ and $x=0.5$ does not have a potential well. Therefore, we can infer that the axial echo signal does not exist. In the BH case, the effective potential does not have a potential well since the perturbed wave propagates through one-way membrane called the event horizon. Therefore, after the big bump coming from the initial Gaussian wave injection, the signal is no longer exist.

On the other hand, with the gravitational echoes, we can also mention the corresponding echo time. This latter quantity describes the time needed for massless particle travel from the center of the star to the photon sphere position. The expression can be written as follows
\begin{equation}
    \tau_{\textrm{echo}} = \int_{0}^{3M} \sqrt{-\frac{g_{rr}}{g_{tt}}} dr,
\end{equation}
where the metric in the integral is the complete set of the solution from the interior until the exterior region. However, with the proposed gravitational wave event, we can infer that the star must have a photon sphere to produce a wave that travels with the characteristic frequency (quasinormal mode). The estimated frequency \cite{Mannarelli:2018pjb} is given by $f_{\textrm{echo}} = 1/2\tau_{\textrm{echo}}.$ In GW170817, the event shows that the stellar can have the frequency around 72 Hz with a $4.2\sigma$ significant level \cite{Mannarelli:2018pjb}. In our case the echo time shifted to the dimensionless form
\begin{equation}
    \tilde{\tau} \equiv \tau_{\textrm{echo}}/\ell,
\end{equation}
with this value at hand, the dimensionless frequency can be written as 
\begin{equation}
    \tilde{f} \equiv f_{\textrm{echo}}\ell.
\end{equation}
It is shown that the gravitational echoes appear only in the condition $x>x_{m}$, and we set $x=0.9$ for $\bar{R}=1.1$ and $\bar{R}=1.5$. We demonstrate, in contrast, that it is possible to achieve a frequency of 72 Hz with the following $\ell$
\begin{eqnarray}
    \ell_{\bar{R}=1.1}&=&1.51\times10^{5} ~\textrm{m}, \\ && \textrm{and}~~~~~\nonumber \\ \ell_{\bar{R}=1.5}&=& 1.33\times10^{5} ~\textrm{m}. 
\end{eqnarray}
As we can see from the above results, the requirement of extremely large $\ell$ values to reproduce the GW170817 frequency makes our anisotropic gravastar scenario less viable.

Regardless of the large predicted value of $\ell$ from potential observational constraints, one can still determine the size of our proposed object. For $x = 0.9$ and $\ell \approx 1.51 \times 10^5\,\text{m}$ ($\bar{R} = 1.1$), using the relation from Eq.~\eqref{eq. alpha def} and restoring natural units via $\mathcal{M} \to G\mathcal{M}/c^2$, we obtain the mass of the object $\mathcal{M}_o$ as
\begin{equation}
    \mathcal{M}_o \approx 115.7 M_\odot,
\end{equation}
where $M_\odot$ is the solar mass. This indicates that our object, if interpreted as the source of gravitational wave echoes from GW170817, would fall within the stellar to intermediate-mass BH range. Nevertheless, this mass can be reduced by choosing a higher value of $x$, which may result in significantly lower echo frequencies compared to those shown in Figs.~\ref{fig:gravechoR11} and \ref{fig:gravechoR15}, and, consequently, a smaller value of $\ell$.

\section{Discussion}
\label{sec. summary}
In this study, we construct an exotic star model inspired by a horizonless RBH solution that is connected to an anisotropic gravastar. The RBH model is modified by adding a correction term to define a radius $R$, making it more suitable for a star solution. The radial metric solution is continuous up to its first derivative, where the modification from Schwarzschild only occur at $r<R$ and it interpolates with the Schwarzschild spacetime in $r\geq R$. As a consequence to the modification, the critical $\alpha$ value is now dependent on $R$, and approaches the standard $\alpha$ in the limit of $R\to\infty$.

We first derive an ansatz for the EoS to simplify the calculations and propose an appropriate function that satisfies the requirements of the anisotropic gravastar. The EoS ansatz shows that, if the radial pressure deviates from the de Sitter EoS ($F(y)\Theta(x-1)\neq0$), there must be a violation of DEC on the transverse pressure once the local compactness of the object is close to forming a horizon. DEC can also be violated with steep transition from negative to positive radial pressure. This implies that this object must be supported by an extreme tension to support the high compactness and/or steep radial pressure transitions.

We provide a phenomenological model example for the radial pressure profile with the deviation proportional to a tangential function. The model includes a de Sitter core that violates the SEC and transitions to a positive-pressure matter region at the outer atmosphere, governed by a polytropic-like EoS. As the configuration approaches the formation of a horizon, the de Sitter core grows, eventually transitioning into a RBH with a pure de Sitter interior ($p=-\epsilon$) once a horizon is formed. By adjusting the values of the parameters, the speed of sound can be kept subluminal in the atmosphere. From the resulting metric solution, we observe a redshift effect in the central core region of the star, which distinguishes it from the metric solution of a standard horizonless RBH where no time delay is present at the center. At the extremal configuration, our results indicate the emergence of a frozen star-like structure on the time geometry ($e^{\nu}$ component), which remains nearly constant and close to zero below the critical point $y_c$, effectively indicating that time freezes in this interior region. We also show that our model can reproduce the thin-shell gravastar model by setting a constant energy density and a step-function pressure transition just beneath the surface radius.

The optical appearance of our star model, surrounded by an accretion disk and assuming that photons do not interact with matter inside the interior region, is distinguishable from BHs, especially with low inclination observation angles close to the axial direction. With an ISCO-profile accretion disk, multiple photon rings are formed, which could potentially be observed with higher resolution EHT. It is shown that there are three major photon rings in the presence of a photon sphere, along with several chaotic minor photon rings between the first and second outer rings. The chaotic minor photon rings are produced by photons with light ring impact parameters. Fewer secondary images are produced in the absence of a photon sphere, and the chaotic minor photon rings are fully resolved, as no light ring trajectories are generated by the spacetime. At high inclination angles, Doppler redshift effects can suppress the appearance of these secondary image intensities, making them more difficult to detect with our finite resolution telescope. Additionally, we check the consistency of our star model and ray-tracing code with the thin-shell gravastar model and find that it is qualitatively comparable to the optical appearance of the thin-shell gravastar as given in Ref.~\cite{Rosa:2024bqv}.

On the other hand, gravitational perturbation is another phenomenological aspect in this paper. We employ the latter method to the unperturbed Einstein field equation with our star model. The master equation leads us to three important aspects, that is effective potential, quasinormal mode, and gravitational echoes. To illustrate the profile, we fix $\bar{R}=1.1$ and $\bar{R}=1.5$. For each of these radii, we use the smooth transition parameter from the horizonless star that has a potential well until it disappears. A profile with $x=x_{m}$ appears with a very small potential well and thus the gravitational wave echo does not exist. Therefore, we can learn that the gravitational wave echo signal begins to appear only when the ultracompact star has a deep and wide potential well for $x>x_{m}$. From the QNM, it also can be inferred that the horizonless ultracompact object characterized by $\bar{R}=1.1$ supports axial perturbations with longer damping timescales than those observed in the model with $\bar{R}=1.5$. Moreover, reproducing the observed frequency of GW170817 requires extremely large values of $\ell$, which may challenge the interpretation of the source as an anisotropic gravastar based on our constructed model.

\section*{Acknowledgments}
BNJ is supported by the Second Century Fund (C2F), Chulalongkorn University, Thailand, and the Research Abroad Scholarship, Chulalongkorn University, Thailand. AS is supported by PUTI grant Universitas Indonesia 2024–2025 No. NKB-380/UN2.RST/HKP.05.00/2024.


\end{document}